\documentstyle[prl,aps,psfig]{revtex}
\def\Bf#1{\mbox{\boldmath{$#1$}}}
\def\bF#1{\mbox{\scriptsize\boldmath{$#1$}}}

\def\Ieq#1#2{{\mathstrut_{#1}^{#2}}}

\def\wt#1{\widetilde{#1}}
\def\wh#1{\widehat{#1}}
\def\ol#1{\overline{#1}}

\def\tJ{$t$-$J$ }
\begin{document} 

\title{
The non-Fermi-liquid nature of the metallic states of the Hubbard 
Hamiltonian } 

\author{\sc Behnam Farid}

\address{Spinoza Institute, Department of Physics and Astronomy,
University of Utrecht,\\
Leuvenlaan 4, 3584 CE Utrecht, The Netherlands
\footnote{Electronic address: B.Farid@phys.uu.nl.} }

\date{Received 13 November 2002 and accepted 25 March 2003}

\maketitle

\begin{abstract}
\leftskip 54.8pt
\rightskip 54.8pt
It has long since been argued that the metallic states of the single-band 
Hubbard Hamiltonian $\wh{\cal H}$ in two spatial dimensions (i.e. for
$d=2$) should be non-Fermi liquid, a possibility that would lead to 
the understanding of the observed anomalous behaviour of the doped 
copper-oxide-based superconducting compounds in their normal metallic 
states. Here we present a formalism which enables us to express, for 
arbitrary $d$, the behaviour of the momentum-distribution function 
${\sf n}_{\sigma}({\Bf k})$ pertaining to uniform metallic ground 
states of $\wh{\cal H}$ close to ${\cal S}_{{\sc f};\sigma}$ (the 
Fermi surface of the fermions with spin index $\sigma$, $\sigma=
\uparrow,\downarrow$) in terms of a small number of constant 
parameters which are bound to satisfy certain inequalities implied 
by the requirement of the stability of the ground state of the 
system. These inequalities restrict the range of variation of 
${\sf n}_{\sigma}({\Bf k})$ for ${\Bf k}$ infinitesimally inside and 
outside the Fermi sea pertaining to fermions with spin index $\sigma$ 
and consequently the range of variation in the zero-temperature limit 
of ${\sf n}_{\sigma}({\Bf k})$ for ${\Bf k}$ on ${\cal S}_{{\sc f};
\sigma}$. On the basis of some available accurate numerical results
for ${\sf n}_{\sigma}({\Bf k})$ pertaining to the Hubbard and the \tJ 
Hamiltonian, we conclude that, at least in the strong-coupling regime, 
the metallic ground states of $\wh{\cal H}$ in $d=2$ {\sl cannot} be 
Fermi-liquid nor can they in general be purely Luttinger or 
marginal Fermi liquids. We further rigorously identify the pseudogap 
phenomenon, or `truncation' of the Fermi surface, clearly observed in 
the normal states of underdoped copper-oxide-based superconductors, as 
corresponding to a line of {\sl resonance} energies (i.e. these 
energies strictly do {\sl not} relate to quasiparticles) located 
{\sl below} the Fermi energy, with a concomitant suppression to zero 
of the jump in ${\sf n}_{\sigma}({\Bf k})$ over the `truncated' 
parts of the Fermi surface. Our analyses make explicit the singular 
significance of the non-interacting energy dispersion 
$\varepsilon_{\bF k}$ underlying $\wh{\cal H}$ in determining the 
low-energy spectral and transport properties of the metallic ground
states of $\wh{\cal H}$.

\end{abstract}

\vspace{1.5cm}
\noindent
\underline{\sf Preprint number: SPIN-2002/01 }
\vspace{1.0cm}

\noindent
\underline{\sf arXiv:cond-mat/0211244~v1~13~Nov~2002} \\
\underline{\sc Philosophical Magazine, 2003, Vol.~83,
No.~24, 2829-2863}
\narrowtext
\twocolumn

\subsection*{\S~1. \sc Introduction}

The unconventional behaviour of the normal states of the doped 
copper-oxide-based high-temperature superconducting compounds (hereafter
referred to as {\sl cuprates}) \cite{OM00,PA01} has necessitated 
reconsideration of the conventional theory of metals \cite{PWA97,GB01}. 
The latter relies on the fundamental premise that in these the low-lying 
single-particle excitations are analogous to those of non-interacting 
metallic systems and that the electron-electron interaction merely 
gives rise to finite renormalization of the low-energy or 
low-temperature response and transport properties of the corresponding 
non-interacting systems. For the last approximately four decades, this 
perspective, which constitutes the phenomenological theory of Landau 
\cite{L57}, has a formal basis on the foundation of the theory of 
many-particle systems, which provides the recipe for the calculation 
of the phenomenological parameters of the Landau theory in terms of 
some material-specific parameters (such as the concentration of the 
carriers) and the fundamental constants of nature \cite{PN66,BP78}.

It is generally considered that the single-band Hubbard Hamiltonian 
$\wh{\cal H}$ for fermions \cite{PWA59,R62,GW63,IKK63,JHI63,JK63} 
confined to a planar square lattice captures the most significant 
aspects of the electronic correlation in the layered cuprate compounds 
\cite{PWA87,PWA88,PWA97} (for a review see \cite{ED94}). In this work 
we expose some exact properties concerning the uniform ground states 
(GSs) and the corresponding low-lying single-particle 
excitations of this Hamiltonian for arbitrary spatial dimensions $d$. 
We derive some novel expressions for ${\sf n}_{\sigma}({\Bf k})$ 
corresponding to ${\Bf k}$ in the close neighbourhood of 
${\cal S}_{{\sc f};\sigma}$ for any conceivable uniform metallic GS 
of $\wh{\cal H}$ in terms of some {\sl constrained} parameters 
deducible from $\varepsilon_{\bF k}$, $\int_{-\infty}^{\mu} 
{\rm d}\varepsilon\; A_{\sigma}({\Bf k};\varepsilon) \equiv \hbar 
{\sf n}_{\sigma}({\Bf k})$ and $\int_{-\infty}^{\mu} 
{\rm d}\varepsilon\;\varepsilon\, A_{\sigma}({\Bf k};\varepsilon)$ 
for ${\Bf k}\to {\cal S}_{{\sc f};\sigma}$, where $A_{\sigma}({\Bf k};
\varepsilon)$ stands for the single-particle spectral function of 
the interacting system and $\mu$ for the `chemical potential' (see 
later). Evidently, since ${\sf n}_{\sigma}({\Bf k})$ itself underlies 
our expressions for ${\sf n}_{\sigma}({\Bf k})$ corresponding to 
${\Bf k} \to {\cal S}_{{\sc f};\sigma}$, these expressions are 
implicit functions of ${\sf n}_{\sigma}({\Bf k})$ for ${\Bf k}
\to {\cal S}_{{\sc f};\sigma}$, implying a closer interdependence 
of the parameters determining the behaviour of ${\sf n}_{\sigma}
({\Bf k})$ for ${\Bf k}\to {\cal S}_{{\sc f};\sigma}$ than is strictly 
relevant for our specific considerations in this work. The latter 
interdependence may be fruitfully incorporated in the design of the 
theoretical schemes that are aimed to the {\sl quantitative} 
calculation of ${\sf n}_{\sigma}({\Bf k})$ and other correlation 
functions pertaining to uniform metallic GSs of $\wh{\cal H}$ 
\cite{BF02}.

On the basis of the above-mentioned expressions for ${\sf n}_{\sigma}
({\Bf k})$, from an available numerical result (obtained within the 
framework of the fluctuation-exchange (FLEX) approximation) concerning 
${\sf n}_{\sigma}({\Bf k})$ corresponding to the Hubbard Hamiltonian 
in $d=2$, with $U/t=8$ and $n_{\sigma}=n_{\bar\sigma}=0.265$ 
\cite{SH91}, we establish that the underlying metallic state
{\sl cannot} be a Fermi-liquid (FL). Here $n_{\sigma}$ stands for the
number of fermions with spin index $\sigma$ per site ($\bar\sigma$
denotes the spin index complementary to $\sigma$ so that, for $\sigma
=\uparrow$, we have $\bar\sigma=\downarrow$), $t$ for the
nearest-neighbour hopping integral and $U$ for the on-site, or
intra-atomic, interaction energy in $\wh{\cal H}$. We arrive at a 
similar conclusion by considering some available accurate numerical 
results concerning ${\sf n}_{\sigma}({\Bf k})$ pertaining to the 
\tJ Hamiltonian for $d=2$, with $J/t=0.4$ and $n_{\sigma}=
n_{\bar\sigma}=0.4$ \cite{PLS98a,PLS98b}, where $J$ stands for the 
exchange integral which in the strong-coupling regime is formally 
equal to $4 t^2/U$ \cite{CSO78,GJR87} (thus $J/t \approx 0.4$ formally 
corresponds to $U/t \approx 10$). 
\footnote{\label{f0}
We should point out that the strong-coupling limit of the Hubbard 
Hamiltonian involves in addition to the \tJ Hamiltonian a three-site 
term, proportional to $t^2/U$. In identifying the \tJ Hamiltonian 
with the strong-coupling limit of the Hubbard Hamiltonian, one 
therefore discards the latter term, a practice which has no 
{\sl a priori} justification away from half-filling.
See \cite{ED94} and \cite{EE96}. }

Our considerations in this work are centred around a {\sl fictitious} 
single-particle Green function ${\cal G}_{\sigma}({\Bf k};\varepsilon)$ 
which is based on the {\sl exact} $N$-particle GS $\vert\Psi_{N;0}
\rangle$ (for simplicity assumed to be uniform) of the single-band 
Hubbard Hamiltonian $\wh{\cal H}$ and two {\sl variational} Ans\"atze 
concerning the $(N\pm 1)$-particle GS {\sl and} excited states 
of $\wh{\cal H}$, $\{ \vert\Psi_{{N\pm 1};s}\rangle \}$. Here $s$ 
stands for a compound index fully characterizing the correlated state 
$\vert\Psi_{M;s}\rangle$. In general, $s=({\Bf k},\alpha)$ where 
${\Bf k}$ is a wave-vector which in our considerations is confined to 
the first Brillouin zone (1BZ) corresponding to the lattice 
$\{ {\Bf R}_j \}$ underlying $\wh{\cal H}$, and $\alpha$ (also a 
compound index) is referred to as the `parameter of degeneracy' 
\cite{KP58,BF01}. The necessity to invoke this additional parameter 
reflects the overcompleteness of the set of single-particle 
excitations of $\wh{\cal H}$ (in fact of {\sl any} interacting 
Hamiltonian) in regard to the underlying single-particle Hilbert 
space \cite{BF01}. In what follows we employ $s=0$ to represent GSs, 
for example as in $\vert\Psi_{M;0}\rangle$, the $M$-particle GS of 
$\wh{\cal H}$.

We demonstrate that insofar as a limited number of characteristics of 
the interacting system are concerned, ${\cal G}_{\sigma}({\Bf k};
\varepsilon)$ is equivalent with the exact single-particle Green 
function $G_{\sigma}({\Bf k};\varepsilon)$. The most relevant of the 
latter characteristics are 

\vspace{1.5mm}
\indent
(i) the GS energy $E_{N;0}$, \\
\indent
(ii) the Fermi energy $\varepsilon_{\sc f}$, \\ 
\indent
(iii) the Fermi surface ${\cal S}_{{\sc f};\sigma}$ and \\ 
\indent
(iv) the GS momentum distribution function ${\sf n}_{\sigma}
({\Bf k})$.

\vspace{1.5mm}
\noindent
On account of these and the relative simplicity of the analytic 
structure of ${\cal G}_{\sigma}({\Bf k};\varepsilon)$, we deduce a 
number of significant facts concerning the GS and the low-lying 
single-particle excitations of the uniform metallic GSs of 
$\wh{\cal H}$. We deduce for instance that the Fermi surface 
${\cal S}_{{\sc f};\sigma}$, $\forall\sigma$, pertaining to one 
such state is a {\sl subset} (not necessarily a {\sl proper} subset) of 
the Fermi surface ${\cal S}_{{\sc f};\sigma}^{(0)}$, $\forall\sigma$, 
that is that determined by the non-interacting energy dispersion 
$\varepsilon_{\bF k}$ and the number $N_{\sigma}$ of particles with 
spin index $\sigma$ corresponding to the interacting $(N_{\sigma}
+N_{\bar\sigma})$-particle uniform GS of $\wh{\cal H}$; leaving aside 
the possibility that the set ${\cal S}_{{\sc f};\sigma}^{(0)}\backslash 
{\cal S}_{{\sc f};\sigma}$ of ${\Bf k}$ points may be non-empty (see 
later), interaction is seen to determine the topology of 
${\cal S}_{{\sc f};\sigma}$ solely through determining the subdivision 
of $N$ into $N_{\sigma}$ and $N_{\bar\sigma}$. In cases where 
${\cal S}_{{\sc f};\sigma}$ is a {\sl proper} subset of 
${\cal S}_{{\sc f};\sigma}^{(0)}$, ${\cal S}_{{\sc f};\sigma}^{(0)} 
\backslash {\cal S}_{{\sc f};\sigma}$ constitutes the {\sl pseudogap} 
region of the putative Fermi surface of the interacting metallic state. 
Our finding that ${\cal S}_{{\sc f};\sigma} \subseteq {\cal S}_{{\sc f};
\sigma}^{(0)}$, directly leads us to conclude that the Luttinger theorem 
\cite{LW60,L60} concerning the `volume' of the interacting Fermi sea 
unreservedly applies to the Fermi sea pertaining to uniform 
metallic GSs of $\wh{\cal H}$. It further leads us to uncover a 
kinematic constraint that in many instances bars partially 
spin-polarized uniform states from being legitimate GSs of the Hubbard 
Hamiltonian. We relegate an extensive discussion of this kinematic
constraint to a future publication; for now however, we only mention 
that the existence of this constraint clarifies the conspicuous absence 
of partially polarized ferromagnetic regions in large portions of the 
zero-temperature phase diagram of the Hubbard Hamiltonian, specifically 
in the strictly tight-binding limit where direct hopping of fermions 
is restricted to their nearest neighbours (see, for instance, 
\cite{MH95,PF97}). We thus mainly ascribe the failure of the Stoner 
criterion in its incorrect prediction of uniform ferromagnetic GSs in 
regions of the latter phase diagram where there is none (see, e.g., 
Fig.~5 in \cite{FOH97}), to the disregard by the Stoner theory of the 
above-mentioned kinematic constraint.

We further deduce that the FL metallic state breaks down at locations 
of the interacting Fermi surface ${\cal S}_{{\sc f};\sigma}$ coinciding 
with saddle points of the non-interacting single-particle energy 
dispersion $\varepsilon_{\bF k}$ and that the two characteristic 
features of the pseudogap region of the putative Fermi surfaces of 
interacting metallic states are inseparably interdependent. To clarify 
the latter aspect, we point out that experimentally pseudogap regions 
of the putative Fermi surfaces of the (underdoped) cuprate compounds 
\cite{AGL96,MDL96,MRN98,AI99} are characterized by: 

\vspace{1.5mm}
\indent
(a) the absence of the quasiparticle peak in the single-particle 
spectral function at the Fermi energy and\\
\indent
(b) the presence of a so-called `leading-edge' peak in the same 
function at a lower energy. 

\vspace{1.5mm}
\noindent
We show that these two aspects are inviolably related.

\vspace{0mm}
\subsection*{\S~2. \sc Preliminaries}

For the single-band Hubbard Hamiltonian \cite{R62,JHI63} corresponding 
to a $d$-dimensional lattice of $N_{\sc l}$ sites we have
\begin{equation}
\label{e1}
\wh{\cal H} = \sum_{i,j} \sum_{\sigma} T_{i,j}\,
{\hat c}^{\dag}_{i;\sigma} {\hat c}_{j;\sigma}
+ \frac{1}{2} U \sum_{i} \sum_{\sigma}
{\hat n}_{i;\sigma} {\hat n}_{i;\bar\sigma},
\end{equation}
where ${\hat c}_{i;\sigma}^{\dag}$, ${\hat c}_{i;\sigma}$ are canonical 
creation and annihilation operators respectively for fermions with spin 
index $\sigma$ at site $i$ corresponding to site vector ${\Bf R}_i$, 
${\hat n}_{i;\sigma} {:=} {\hat c}_{i;\sigma}^{\dag} {\hat c}_{i;\sigma}$
is a partial site number operator, $U$ the on-site interaction energy 
(in this work assumed to be positive) and 
\begin{equation}
\label{e2}
T_{i,j} = \frac{1}{N_{\sc l}} \sum_{\bF k} \varepsilon_{\bF k}\, 
{\rm e}^{ i {\bF k}\cdot ({\bF R}_i - {\bF R}_j) } 
\end{equation}
a hopping matrix element corresponding to the non-interacting 
(spin-degenerate) energy dispersion $\varepsilon_{\bF k}$. For 
simplicity, in this work we only consider the GSs of $\wh{\cal H}$ 
which preserve the full translational symmetry of the underlying Bravais 
lattice $\{ {\Bf R}_i \| i=1,\dots, N_{\sc l} \}$ so that both in 
Eq.~(\ref{e2}) above and later, $\sum_{\bF k}$ signifies a sum over 
$N_{\sc l}$ points in the first Brillouin zone (1BZ) corresponding 
to the lattice $\{ {\Bf R}_i \}$. It is common practice to choose 
the origin of the energy to coincide with $T_{i,i} = N_{\sc l}^{-1} 
\sum_{\bF k} \varepsilon_{\bF k}$. By doing so, in the strict 
tight-binding limit, $T_{i,j}$ is non-zero only when ${\Bf R}_i$ and 
${\Bf R}_j$ are nearest neighbours, in which case it is identified 
with $-t < 0$. None of our following considerations relies on a 
strict tight-binding approximation to $\{ T_{i,j} \}$. Since the partial 
number operators $\wh{N}_{\sigma} = \sum_i {\hat c}_{i;\sigma}^{\dag} 
{\hat c}_{i;\sigma}$, $\forall\sigma$, commute with $\wh{\cal H}$, 
the uniform states of $\wh{\cal H}$ can be characterized by 
$n_{\sigma} {:=} N_{\sigma}/N_{\sc l} \equiv n - n_{\bar\sigma}$,
where $n {:=} N/N_{\sc l}$ with $N = \sum_{\sigma} N_{\sigma}$.

\subsection*{\S~3. \sc The single-particle Green function and some 
exact sum rules}

We seek to investigate the single-particle properties of the many-body 
Hamiltonian $\wh{\cal H}$ as embodied by the single-particle Green 
function $G_{\sigma}(\varepsilon)$ \cite{FW71}, with $\varepsilon$ 
the external energy parameter. For the coordinate representation of 
$\wt{G}_{\sigma}(z)$, the analytically continued $G_{\sigma}
(\varepsilon)$ into the physical Riemann sheet of the complex $z$ 
plane (see later), we have
\begin{equation}
\label{e3}
\wt{G}_{\sigma}({\Bf r},{\Bf r}';z)
= \sum_{\bF k} \wt{G}_{\sigma}({\Bf k};z)\,
\psi_{\bF k}({\Bf r}) \psi_{\bF k}^*({\Bf r}'),
\end{equation}
where $\psi_{\bF k}({\Bf r})$ is the normalized Bloch function 
corresponding to $\varepsilon_{\bF k}$ and
\begin{equation}
\label{e4}
\wt{G}_{\sigma}({\Bf k};z) = \hbar \sum_s
\frac{ \vert {\sf\bar f}_{s;\sigma}({\Bf k}) \vert^2}
{z - \varepsilon_{s;\sigma} }.
\end{equation}
The physical Green function $G_{\sigma}({\Bf k};\varepsilon)$, with 
$\varepsilon$ real, is obtained through $G_{\sigma}({\Bf k};\varepsilon) 
= \lim_{\eta\downarrow 0} \wt{G}_{\sigma}({\Bf k};\varepsilon\pm i\eta)$, 
$\varepsilon\, \Ieq<>\, \mu$, where $\mu$ stands for the `chemical 
potential' to be specified below. With $\{ \vert\Psi_{N_{\sigma} \pm 1, 
N_{\bar\sigma};s} \rangle \,\|\, s\}$ the complete set of $(N_{\sigma}
\pm 1 + N_{\bar\sigma})$-particle eigenstates of $\wh{\cal H}$ (the 
charge-neutral system accommodates $(N_{\sigma} + N_{\bar\sigma}) 
\equiv N$ particles) and the corresponding eigenenergies
$\{ E_{N_{\sigma} \pm 1, N_{\bar\sigma};s} \,\|\, s\}$, the Lehmann 
amplitudes $\{ {\sf\bar f}_{s;\sigma}({\Bf k}) \}$ and the associated
single-particle energies $\{ \varepsilon_{s;\sigma} \}$ are defined 
as follows:
\begin{eqnarray}
\label{e5}
{\sf\bar f}_{s;\sigma}({\Bf k})
{:=} \left\{ \begin{array}{ll}
\langle\Psi_{N_{\sigma}-1,N_{\bar\sigma};s}\vert
{\hat c}_{{\bF k};\sigma} \vert\Psi_{N;0}\rangle,
&\;\; \varepsilon_{s;\sigma} < \mu, \\ \\
\langle\Psi_{N;0}\vert {\hat c}_{{\bF k};\sigma}
\vert\Psi_{N_{\sigma}+1,N_{\bar\sigma};s}\rangle,
&\;\; \varepsilon_{s;\sigma} > \mu,
\end{array} \right.
\end{eqnarray}
\begin{eqnarray}
\label{e6}
\varepsilon_{s;\sigma} {:=} \left\{ \begin{array}{ll}
E_{N;0} - E_{N_{\sigma}-1,N_{\bar\sigma};s},
&\;\;\; \varepsilon_{s;\sigma} < \mu, \\ \\
E_{N_{\sigma}+1,N_{\bar\sigma};s} - E_{N;0},
&\;\;\; \varepsilon_{s;\sigma} > \mu,
\end{array} \right.
\end{eqnarray}
where ${\hat c}_{{\bF k};\sigma} \equiv N_{\sc l}^{-1/2} \sum_j
{\hat c}_{j;\sigma}\,\exp(-i {\Bf k}\cdot {\Bf R}_j)$, $\mu$ is 
defined in Eq.~(\ref{e8}) below, and $\vert\Psi_{N;0}\rangle$ and 
$E_{N;0}$ are the short-hand notations for $\vert\Psi_{N_{\sigma},
N_{\bar\sigma};0}\rangle$ and $E_{N_{\sigma},N_{\bar\sigma};0}$ 
respectively. 

Our notation in Eq.~(\ref{e6}) is indicative of $\{ \varepsilon_{s;
\sigma} \}$ consisting of two {\sl disjoint} subsets, an aspect which 
has its root in our assumption concerning the stability of the 
$(N_{\sigma}+N_{\bar\sigma})$-particle GS of $\wh{\cal H}$.
By defining
\begin{eqnarray}
\label{e7}
\mu_{N;\sigma}^- &{:=}& E_{N;0}-E_{N_{\sigma}-1,N_{\bar\sigma};0},
\nonumber\\
\mu_{N;\sigma}^+ &{:=}& E_{N_{\sigma}+1,N_{\bar\sigma};0}-E_{N;0},
\end{eqnarray}
the disjoint nature of the above-mentioned two sets of excitation
energies can be expressed as follows
\begin{equation}
\label{e8}
\mu_{N;\sigma}^- < \mu < \mu_{N;\sigma}^+,
\;\;\; \forall \sigma.
\end{equation}
For non-metallic states, $\mu_{N;\sigma}^+ - \mu_{N;\sigma}^-$
is finite, equal to the fundamental gap in the single-particle
excitation spectrum of the system. For metallic (thus microscopic)
GSs, $\mu_{N;\sigma}^+ - \mu_{N;\sigma}^-$ is infinitesimally small, 
scaling like $1/N^p$ with $p > 0$; however it is {\sl non-vanishing}. 
For these states, the Fermi energy $\varepsilon_{\sc f}$ is formally 
equal to $\mu_{N;\sigma}^-$ (in which case the subscript $\sigma$ 
is redundant when $N_{\sigma},N_{\bar\sigma}\not=0$; it is {\sl not} 
redundant when $N_{\bar\sigma}=0$) and its undue identification with 
$\mu$ (which lies in an {\sl open} interval) can give rise to 
erroneous or ill-defined results. To appreciate this aspect,
consider the single-particle spectral function
\begin{eqnarray}
\label{e9}
A_{\sigma}({\Bf k};\varepsilon) &{:=}&
\frac{1}{2\pi i}\left. \big[
\wt{G}_{\sigma}({\Bf k};\varepsilon-i\eta) -
\wt{G}_{\sigma}({\Bf k};\varepsilon+i\eta) \big]
\right|_{\eta\downarrow 0} \nonumber\\
&\equiv& 
\hbar \sum_s \vert {\sf\bar f}_{s;\sigma}({\Bf k})\vert^2\,
\delta(\varepsilon-\varepsilon_{s;\sigma}),
\end{eqnarray}
for which we have
\begin{eqnarray}
\label{e10}
&&\frac{1}{\hbar} \int_{-\infty}^{\mu}
{\rm d}\varepsilon\; A_{\sigma}({\Bf k};\varepsilon)
= {\sf n}_{\sigma}({\Bf k}), \\
\label{e11}
&&\frac{1}{\hbar} \int_{\mu}^{\infty}
{\rm d}\varepsilon\; A_{\sigma}({\Bf k};\varepsilon)
= 1 - {\sf n}_{\sigma}({\Bf k}).
\end{eqnarray}
For some metallic states (satisfying condition (A) introduced in 
\S~11.2.2 below), including FLs, ${\sf n}_{\sigma} ({\Bf k})$ 
undergoes a discontinuous change upon transposing ${\Bf k}$ from 
infinitesimally inside to infinitesimally outside the Fermi sea; 
the amount of discontinuity is by a Migdal theorem \cite{ABM57,L60} 
equal to the weight of a $\delta$-function contribution to 
$A_{\sigma}({\Bf k};\varepsilon)$ at $\varepsilon=\varepsilon_{\sc f}$, 
the Fermi energy. From Eq.~(\ref{e10}) it is observed that, for this 
coherent contribution to $A_{\sigma}({\Bf k};\varepsilon)$ to have 
an impact on the behaviour of ${\sf n}_{\sigma}({\Bf k})$ for ${\Bf k} 
\to {\cal S}_{{\sc f};\sigma}$, it is required that $\varepsilon_{\sc f} 
\equiv \mu_{N;\sigma}^- < \mu$. Similarly, from Eq.~(\ref{e11}) we 
observe that there must exist a comparable coherent contribution to 
$A_{\sigma}({\Bf k};\varepsilon)$ at an $\varepsilon$ inside the 
interval $(\mu,\infty)$; for ${\Bf k} \to {\cal S}_{{\sc f};\sigma}$, 
the required $\delta$-function contribution is located at $\varepsilon
=\mu_{N;\sigma}^+ > \mu$ (see Eq.~(\ref{e8})). The formal mode of 
expressing these observations is firstly that, the singular points of 
$A_{\sigma}({\Bf k};\varepsilon)$ coincide with the solutions of
\begin{equation}
\label{e12}
\varepsilon_{\bF k} + 
\hbar\Sigma_{\sigma}({\Bf k};\varepsilon) = \varepsilon,
\end{equation}
where $\Sigma_{\sigma}({\Bf k};\varepsilon)$ stands for the self-energy
and secondly, that this equation has one solution at $({\Bf k},\varepsilon) 
= ({\Bf k}_{{\sc f};\sigma}^-,\mu_{N;\sigma}^-)$ and one solution at 
$({\Bf k},\varepsilon) = ({\Bf k}_{{\sc f};\sigma}^+, \mu_{N;\sigma}^+)$, 
where ${\Bf k}_{{\sc f};\sigma}^{-}$ and ${\Bf k}_{{\sc f};\sigma}^{+}$ 
are respectively infinitesimally inside and outside the Fermi sea 
pertaining to fermions with spin index $\sigma$ in the vicinity of 
${\Bf k}_{{\sc f};\sigma} \in {\cal S}_{{\sc f};\sigma}$. The necessity 
for the introduction of two vectors ${\Bf k}_{{\sc f};\sigma}^-$ and 
${\Bf k}_{{\sc f};\sigma}^+$ arises from the fact that in principle 
(see \S~3.4 in \cite{BF01}), Eq.~(\ref{e12}), which corresponds to a 
{\sl uniform} GS, cannot possess more than one solution at any 
given ${\Bf k}$; in the case of the Hubbard Hamiltonian, where the 
momentum space is restricted to the 1BZ corresponding to the underlying 
lattice, all functions of ${\Bf k}$ are periodic with the periodicity 
of the 1BZ, so that through Umklapp processes the latter principle is 
not of general validity (see below). 

For an arbitrary ${\Bf k}$, Eq.~(\ref{e12}) has either no solution 
or it has real-valued solution(s) (see above) \cite{BF01}. 
From the above considerations we deduce that a singular behaviour 
(not necessarily a discontinuity) in ${\sf n}_{\sigma}({\Bf k})$ 
at ${\Bf k}={\Bf k}_0 \not={\Bf k}_{{\sc f};\sigma}^{\mp}$, with 
${\Bf k}_{{\sc f};\sigma} \in {\cal S}_{{\sc f};\sigma}$, implies 
the existence of at least two (real-valued) solutions 
$\varepsilon_{{\bF k}_0;\sigma}^{(i)}$, $i=1,2$, which do not
necessarily correspond to quasiparticle excitations and for which 
$\varepsilon_{{\bF k}_0;\sigma}^{(1)} < \mu_{N;\sigma}^{-}$ and
$\varepsilon_{{\bF k}_0;\sigma}^{(2)} > \mu_{N;\sigma}^+$ hold. 
The validity of the latter inequalities is readily verified by 
considering Eqs.~(\ref{e10}) and (\ref{e11}) along the lines presented 
above. The observation of singular behaviour in ${\sf n}_{\sigma}(k)$
pertaining to the GS of the Hubbard Hamiltonian for $d=1$ in the
limit of $U\to\infty$ at $k$ different from $\pm k_{{\sc f};\sigma}$
(for instance at $k=\pm 3 k_{{\sc f};\sigma}$ in the case corresponding 
to  $n_{\sigma}=n_{\bar\sigma}=1/4$) \cite{OS90,OSS91} implies that 
at these $k$ the corresponding single-particle spectral function must 
possess at least two pronounced peaks, one below $\mu_{N;\sigma}^-$ 
and one above $\mu_{N;\sigma}^+$. 

Making use of the anticommutation relations for the canonical fermion 
operators ${\hat c}_{{\bF k};\sigma}^{\dag}$ and ${\hat c}_{{\bF k};
\sigma}$, it can be readily shown that
\begin{equation}
\label{e13}
\sum_s {\sf\bar f}_{s;\sigma}^*({\Bf k})
{\sf\bar f}_{s;\sigma}({\Bf k}') = \delta_{{\bF k},{\bF k}'},\;\;
{\Bf k}, {\Bf k}' \in \mbox{\rm 1BZ};
\end{equation}
however 
\begin{equation}
\label{e14}
\sum_{\bF k} {\sf\bar f}_{s;\sigma}^*({\Bf k})
{\sf\bar f}_{s';\sigma}({\Bf k}) \not= \delta_{s,s'}
\;\;\; \mbox{\rm for}\;\;\; U\not=0. 
\end{equation}
The latter is a manifestation of the aforementioned over-completeness 
of the set $\{ {\sf\bar f}_{s;\sigma}({\Bf k})\| s\}$ \cite{BF01}; one, 
however, has (for $s = ({\Bf k},\alpha)$ see above) 
\begin{equation}
\label{e15}
\left. {\sf\bar f}_{s;\sigma}({\Bf k}')\right|_{s=({\bF k},\alpha)}
\to \delta_{{\bF k},{\bF k}'}\;\;\; \mbox{\rm for}\;\;\; U\to 0. 
\end{equation}

For our later considerations it is necessary to be aware of the 
following sum rules:
\begin{equation}
\label{e16}
\sum_{s}^{\Ieq<>}
{\sf\bar f}_{s;\sigma}^*({\Bf k})
{\sf\bar f}_{s;\sigma}({\Bf k}') = 
\nu_{\sigma}^{\Ieq<>}({\Bf k})\, \delta_{{\bF k},{\bF k}'},\;\;\;
{\Bf k}, {\Bf k}' \in \mbox{\rm 1BZ},
\end{equation}
where $\sum_s^{\Ieq<>}$ stands for the sum over {\sl all} $s$
for which $\varepsilon_{s;\sigma} \Ieq<> \mu$, and
\begin{eqnarray}
\label{e17}
\nu_{\sigma}^{\Ieq><}({\Bf k})
{:=} \left\{ \begin{array}{l}
{\sf n}_{\sigma}({\Bf k}), \\
1-{\sf n}_{\sigma}({\Bf k}).
\end{array} \right.
\end{eqnarray}
The validity of the results in Eq.~(\ref{e16}) is readily verified
by the knowledge that $\wh{\cal H}$ commutes with the total momentum 
operator $\wh{\Bf P} {:=} \hbar \sum_{{\bF k},\sigma} {\Bf k}\, 
{\hat c}_{{\bF k};\sigma}^{\dag} {\hat c}_{{\bF k};\sigma}$ and that
$\wh{\Bf P} \vert\Psi_{N;0}\rangle = 0$. 

\subsection*{\S~4. \sc Fermi sea, Fermi surface, a Luttinger 
theorem and the pseudogap}

We define the Fermi sea FS$_{\sigma}$ for the fermions with spin 
index $\sigma$ as follows:
\begin{equation}
\label{e18}
{\rm FS}_{\sigma} {:=}
\big\{ {\Bf k} \,\|\, 1/G_{\sigma}({\Bf k};\mu) > 0 \big\} \equiv
\big\{ {\Bf k} \,\|\, G_{\sigma}({\Bf k};\mu) > 0 \big\},
\end{equation}
which can be shown to conform with the definition given by Galitskii 
and Migdal \cite{GM58} and Luttinger \cite{L60} (see Eqs.~(6) and (94) 
in the latter work). Note that, in view of Eqs.~(\ref{e8}) and 
(\ref{e9}), $A_{\sigma}({\Bf k};\mu) \equiv 0$, $\forall {\Bf k}$, so 
that ${\rm Im}[G_{\sigma}({\Bf k};\mu)] \equiv 0$, $\forall {\Bf k}$. 
One readily verifies that the definition in Eq.~(\ref{e18}) identically 
reproduces the definition for the non-interacting Fermi sea 
FS$_{\sigma}^{(0)} {:=} \{ {\Bf k}\, \|\, \epsilon_{{\bF k};\sigma}
\le \varepsilon_{\sc f}^{(0)} \}$ pertaining to the GS of the 
non-interacting Hamiltonian
\begin{equation}
\label{e19}
\wh{\cal H}_0 {:=} \sum_{{\bF k},\sigma}
\epsilon_{{\bF k};\sigma}\,
{\hat c}_{{\bF k};\sigma}^{\dag}
{\hat c}_{{\bF k};\sigma},
\end{equation}
where
\begin{equation}
\label{e20}
\epsilon_{{\bF k};\sigma} {:=} \varepsilon_{\bF k}
+ \epsilon_{\sigma},
\end{equation}
in which the constants $\{ \epsilon_{\sigma} \}$ are chosen such that
\begin{equation}
\label{e21}
\sum_{\bF k} \Theta(\varepsilon_{\sc f}^{(0)} -
\varepsilon_{\bF k} - \epsilon_{\sigma}) = N_{\sigma},\;\;\;
\sum_{\sigma} N_{\sigma} = N;
\end{equation}
in view of our later considerations, we assume that $\{N_{\sigma}\}$ 
corresponds to the {\sl exact} $N$-particle uniform GS of 
$\wh{\cal H}$.

We further define
\begin{equation}
\label{e22}
\ol{\rm FS}_{\sigma} {:=} 
{\rm 1BZ}\backslash {\rm FS}_{\sigma} \equiv
\{ {\Bf k}\, \| \, 1/G_{\sigma}({\Bf k};\mu) < 0 \}.
\end{equation}
With
\begin{equation}
\label{e23}
{\cal S}_{{\sc f};\sigma} {:=}
\{ {\Bf k}\, \| \, 1/G_{\sigma}({\Bf k};\varepsilon_{\sc f}) = 0 \}
\end{equation}
the interacting Fermi surface, it follows that ${\cal S}_{{\sc f};
\sigma} \subset {\rm FS}_{\sigma}$ (for this reason, in the following
we consider ${\Bf k}_{{\sc f};\sigma}^-$ and ${\Bf k}_{{\sc f};\sigma}$ 
as being interchangeable); in defining ${\cal S}_{{\sc f};\sigma}$ it 
has been assumed that $\mu_{N;\sigma}^+ - \mu_{N;\sigma}^-$ scales 
like $1/N^p$, where $p > 0$ and $N\to\infty$, taking into account the 
relationship $\varepsilon_{\sc f} = \mu_{N;\sigma}^-$. It is directly 
verified that the definition in Eq.~(\ref{e23}) reproduces the 
definition for the non-interacting Fermi surface, namely 
${\cal S}_{{\sc f};\sigma}^{(0)} {:=} \{ {\Bf k} \, \| \, 
\epsilon_{{\bF k};\sigma} = \varepsilon_{\sc f}^{(0)}\}$. We should 
point out that ${\cal S}_{{\sc f};\sigma}$ as defined in 
Eq.~(\ref{e23}), as well as its non-interacting counterpart 
${\cal S}_{{\sc f};\sigma}^{(0)}$, need {\sl not} be 
simply-connected, as evidenced by the fact that for $d=1$, for 
instance, neither ${\cal S}_{{\sc f};\sigma}$ nor ${\cal S}_{{\sc f};
\sigma}^{(0)}$ is simply-connected. However, even for cases where 
${\cal S}_{{\sc f};\sigma}^{(0)}$ and ${\cal S}_{{\sc f};\sigma}$ 
are {\sl not} simply-connected sets (either jointly or separately), 
the number of ${\Bf k}$ points inside FS$_{\sigma}^{(0)}$ and 
FS$_{\sigma}$ (the `volume' of the Fermi sea) is well-defined. 
\footnote{\label{f1}
Unless the contexts imply otherwise, our statements here are 
based on the assumption that the systems under consideration are 
in the thermodynamic limit so that, for the purpose of counting 
the number of ${\Bf k}$ points, one can assume a continuous 
(and uniform) distribution of these and thus can disregard sets of 
measure zero when applicable. } 
This is in fact even the case for non-metallic GSs for which 
${\cal S}_{{\sc f};\sigma}$ is an empty set. An aspect of special 
importance, both generally and particularly for our considerations 
in this work, is the validity of a Luttinger theorem 
\cite{L60,LW60} according to which the number of ${\Bf k}$ points 
interior to FS$_{\sigma}$ is equal to $N_{\sigma}$. The non-triviality 
of the Luttinger theorem at issue therefore rests on the definition for 
FS$_{\sigma}$ in Eq.~(\ref{e18}) for {\sl interacting} systems, for 
which {\sl no} orthonormal set of single-particle wavefunctions can be 
associated with the single-particle excitations (see Eq.~(\ref{e14}) 
above). As an essential ingredient of the proof of the Luttinger theorem 
at issue is dependent on a term-by-term analysis of the perturbation 
series for $\Sigma_{\sigma}({\Bf k};\varepsilon)$, it has been variously 
suggested that the Luttinger theorem should {\sl not} be of general 
validity. In this work we demonstrate that, for the uniform metallic 
GSs of $\wh{\cal H}$, 
\begin{equation}
\label{e24}
{\cal S}_{{\sc f};\sigma} \subseteq
{\cal S}_{{\sc f};\sigma}^{(0)}.
\end{equation}
It follows that, even in cases where ${\cal S}_{{\sc f};\sigma} 
\subset {\cal S}_{{\sc f};\sigma}^{(0)}$, the number of ${\Bf k}$ 
points in the interior of FS$_{\sigma}$ is equal to that of 
FS$_{\sigma}^{(0)}$, establishing the validity of the Luttinger 
theorem in the case at hand. The finding to the contrary by 
Schmalian, {\sl et al.} \cite{SLGB96}, is a consequence of 
$\Gamma_{\bF k}$ (following the notation used in \cite{SLGB96}, equal to 
$-{\rm Im}\Sigma_{\bF k}(\omega=0)$), as calculated by these workers,
not to vanish in the zero-temperature limit, in strict violation 
of the property $A_{\sigma}({\Bf k};\mu) \equiv 0 \iff 
{\rm Im}[\Sigma_{\sigma}({\Bf k};\mu)]\equiv 0$, $\forall {\Bf k}$, 
which is a direct consequence of the results in Eqs.~(\ref{e8}) and 
(\ref{e9}) (the second expression) above; inspection of the 
proof of the Luttinger theorem at issue \cite{L60,LW60} makes evident 
that without ${\rm Im}[\Sigma_{\sigma}({\Bf k};\mu)]\equiv 0$, $\forall 
{\Bf k}$, with $\mu$ in the {\it open} interval indicated in 
Eq.~(\ref{e8}) above, the counting device employed by Luttinger 
unequivocally fails.

The result in Eq.~(\ref{e24}) is in conformity with our observation 
\cite{BF02}, to be briefly discussed below (see Eq.~(\ref{e68}) below 
and the subsequent text), that ${\sf n}_{\sigma}({\Bf k})$ pertaining 
to the uniform metallic GSs of $\wh{\cal H}$ is singular (not 
necessarily discontinuous) for all ${\Bf k} \in {\cal S}_{{\sc f};
\sigma}^{(0)}$. In this connection, we note in passing that, on the 
basis of the arguments leading to the latter observation, we \cite{BF02} 
succeed in deducing {\sl all} singularities in ${\sf n}_{\sigma}(k)$ 
pertaining to the GS of the Hubbard Hamiltonian in $d=1$ in the limit 
of infinite $U$, as calculated by Ogata and Shiba \cite{OS90} and Ogata, 
{\sl et al.} \cite{OSS91} (the latter work considers the effects of an 
externally applied magnetic field) on the basis of the Bethe Ansatz
solution for the GS of the Hubbard Hamiltonian. The fact that 
${\sf n}_{\sigma}({\Bf k})$ is singular for {\sl all} ${\Bf k}\in 
{\cal S}_{{\sc f};\sigma}^{(0)}$, even in cases where 
${\cal S}_{{\sc f};\sigma} \subset {\cal S}_{{\sc f};\sigma}^{(0)}$ 
(a {\sl proper} subset) implies that, although for ${\Bf k} \in 
{\cal S}_{{\sc f};\sigma}^{(0)} \backslash {\cal S}_{{\sc f};\sigma}$ 
Eq.~(\ref{e12}) is {\sl not} satisfied for $\varepsilon 
=\varepsilon_{\sc f}$, there must exist an energy $\varepsilon_0$, with 
$\varepsilon_0 < \varepsilon_{\sc f}$, for which this equation is 
satisfied (see Eqs.~(\ref{e10}) and (\ref{e11}) above and the subsequent 
text). Thus the region ${\cal S}_{{\sc f};\sigma}^{(0)} \backslash 
{\cal S}_{{\sc f};\sigma}$, when non-empty, can be identified with 
the pseudogap region, which has been detected, amongst others 
\cite{TS99}, in the experimental photoemission spectra concerning the 
cuprate compounds Bi$_2$Sr$_2$CaCuO$_{8+\delta}$ \cite{AGL96,MDL96,MRN98} 
and La$_{2-x}$Sr$_{x}$CuO$_4$ \cite{AI99} in the underdoped region (for 
a recent review of the photoemission experiments concerning cuprates 
see \cite{DSH02}). Theoretically, these observations have been most 
recently associated with the formation of ``charge and spin gap'' and 
``truncation of the Fermi surface'' into disconnected `arcs' (in $d=2$) 
in the neighbourhoods of the saddle-points of $\varepsilon_{\bF k}$, 
and attributed to ``Umklapp scattering processes across the Fermi 
surface'' \cite{HSFR01}; inspired by the physics of the $M$-leg ladder 
compounds \cite{DR96,DRS92,WS97,BMF96,LBMF97,RHSZ97,LHR00}, the present
authors have identified the corresponding metallic state as being 
an `insulating spin liquid' (ISL) \cite{FRS98,FR98} (see also
\cite{WL96,PAL99,PAL02,CMV97,CMV99,CLMN01}). We relegate a discussion 
of the physical state associated with the pseudogap phenomenon to 
a future publication.

Calculations of ${\cal S}_{{\sc f};\sigma}$ based on the second-order 
perturbation expansion of $\Sigma_{\sigma}({\Bf k};\varepsilon)$ in 
terms of the non-interacting Green function $G_{\sigma}^{(0)}$
indicate ${\cal S}_{{\sc f};\sigma} \not\subseteq {\cal S}_{{\sc f};
\sigma}^{(0)}$ \cite{SG88,ZEG96,HM97,YY99}. By construction,
in these calculations FS$_{\sigma}$ contains the same number of 
${\Bf k}$ points as FS$_{\sigma}^{(0)}$ (in \cite{HM97}, for 
instance, by choosing the second-order change in $\mu$ to be given 
by a Fermi-surface average, the number of ${\Bf k}$ points inside 
the calculated FS$_{\sigma}$ is equal to that inside 
FS$_{\sigma}^{(0)}$ to third order in $U$) so that within the 
frameworks of these calculations the question with regard to the 
validity or otherwise of the Luttinger theorem has {\sl not} been 
considered. The essential shortcoming of the perturbative 
determination of ${\cal S}_{{\sc f};\sigma}$ along the lines 
followed by \cite{SG88,ZEG96,HM97,YY99} becomes evident by realizing 
that one of the two fundamental conditions for ${\Bf k}$ to be located 
on ${\cal S}_{{\sc f};\sigma}$ is that ${\rm Im}[\wt{\Sigma}_{\sigma}
({\Bf k};\varepsilon_{{\bF k};\sigma})] = 0$, where 
$\varepsilon_{{\bF k};\sigma}=\varepsilon_{\bF k} + 
\hbar\wt{\Sigma}_{\sigma}({\Bf k};\varepsilon_{{\bF k};\sigma})$ 
({\it cf}. Eq.~(\ref{e12})), a condition {\sl not} taken account 
of in the cited references; this occurs because, for FLs (note this 
aspect) to second order in $U$, ${\rm Im}[\Sigma_{\sigma}({\Bf k};
\varepsilon)]$ does not feature in the equation for 
${\cal S}_{{\sc f};\sigma}$. Consequently, and as can also be 
explicitly verified, the ${\cal S}_{{\sc f};\sigma}$ obtained by 
\cite{SG88,ZEG96,HM97,YY99} does not satisfy one of the 
necessary conditions defining ${\cal S}_{{\sc f};\sigma}$. The 
gravity of violating this condition can be appreciated by realizing 
the fact that an inappropriate {\sl sign} (to be contrasted with 
magnitude) assigned to ${\rm Im}[\wt{\Sigma}_{\sigma}({\Bf k};
\varepsilon_{{\bF k};\sigma})]$, following an incorrect subdivision 
of the momentum space into interior and exterior of the Fermi sea, 
amounts to the instability of the GS of the system under consideration. 
It follows that ${\cal S}_{{\sc f};\sigma}$ is far more tightly 
constrained than can be determined to second order in $U$ (more
generally, to any {\sl finite} order in $U$). 

\subsection*{\S~5. \sc Formal considerations}

We are now in a position to introduce our formalism that enables us 
to obtain exact results concerning the properties (i) - (iv)
enumerated at the outset of this paper. To this end, we introduce 
the following substitutions:
\begin{eqnarray}
\label{e25}
&& s \rightharpoonup {\tilde{\Bf k}}, \;\;\;\;
{\sf\bar f}_{s;\sigma}({\Bf k}) \rightharpoonup
{\sf\bar f}_{{\tilde{\bF k}};\sigma}({\Bf k}), \;\;\;\;
\varepsilon_{s;\sigma} \rightharpoonup
\varepsilon_{{\tilde{\bF k}};\sigma}, \nonumber\\
&& \sum_s^{<} \rightharpoonup
\sum_{{\tilde{\bF k}}\in {\rm FS}_{\sigma} }, \;\;\;\;
\sum_s^{>} \rightharpoonup
\sum_{{\tilde{\bF k}}\in \overline{\rm FS}_{\sigma} },
\end{eqnarray}
where, in spite of suppressing the `parameter of degeneracy' $\alpha$ 
associated with $s$ (see above), FS$_{\sigma}$ and 
$\overline{\rm FS}_{\sigma}$ are the same as in the exact theory. 

\begin{figure}[t!]
\protect
\centerline{
\psfig{figure=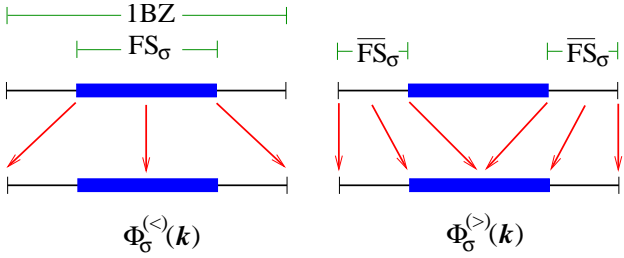,width=3.25in} }
\vskip 5pt
\caption{\label{fi1} \sf
Schematic representation of the one-to-one
mappings defined in Eqs.~(\protect\ref{e26}) and 
(\protect\ref{e27}). }
\end{figure}

Let now ${\Bf\Phi}_{\sigma}^{(<)}({\Bf k})$ and 
${\Bf\Phi}_{\sigma}^{(>)} ({\Bf k})$ be two one-to-one mappings 
according to (see Fig.~\ref{fi1})
\begin{eqnarray}
\label{e26}
&&{\Bf\Phi}_{\sigma}^{(<)}({\Bf k}): {\rm FS}_{\sigma} \mapsto
\mbox{\rm 1BZ},\\
\label{e27}
&&{\Bf\Phi}_{\sigma}^{(>)}({\Bf k}): \overline{\rm FS}_{\sigma} \mapsto
\mbox{\rm 1BZ}.
\end{eqnarray}
These mappings, which are functions of $U$, are required to approach 
the following {\sl identity} mappings 
\begin{eqnarray}
\label{e28}
&&{\Bf\Phi}_{0;\sigma}^{(<)}({\Bf k}):
{\rm FS}_{\sigma}^{(0)} \mapsto {\rm FS}_{\sigma}^{(0)},\\
\label{e29}
&&{\Bf\Phi}_{0;\sigma}^{(>)}({\Bf k}):
\overline{\rm FS}_{\sigma}^{(0)} \mapsto
\overline{\rm FS}_{\sigma}^{(0)},
\end{eqnarray}
for $\wh{\cal H}$ approaching the non-interacting Hamiltonian
$\wh{\cal H}_0$ as defined in Eq.~(\ref{e19}). For 
weakly interacting systems, one therefore expects 
${\Bf\Phi}_{\sigma}^{(<)}({\Bf k})$ and ${\Bf\Phi}_{\sigma}^{(>)}
({\Bf k})$ to resemble closely the identity mappings ${\Bf\Phi}_{0;
\sigma}^{(<)}({\Bf k})$ and ${\Bf\Phi}_{0;\sigma}^{(>)}({\Bf k})$,
specifically for ${\Bf k}$ deep in the interiors of FS$_{\sigma}$ 
and $\overline{\rm FS}_{\sigma}$ respectively. We point out that 
${\Bf\Phi}_{0;\sigma}^{(\Ieq><)}({\Bf k})$ do {\sl not} play any 
explicit role in our following considerations. 

For ${\sf\bar f}_{{\tilde{\bF k}};\sigma}({\Bf k})$ introduced in
Eq.~(\ref{e25}), we now make the following choice:
\begin{eqnarray}
\label{e30}
{\sf\bar f}_{{\tilde{\bF k}};\sigma}({\Bf k})
{:=} \left\{ \begin{array}{ll}
\sqrt{\nu_{\sigma}^{<}({\Bf k})}\,
\delta_{{\bF\Phi}_{\sigma}^{(<)}({\tilde{\bF k}}),{\bF k}},
&\;\; {\tilde{\Bf k}} \in {\rm FS}_{\sigma},\; {\Bf k} \in
\mbox{\rm 1BZ}, \\ \nonumber \\
\sqrt{\nu_{\sigma}^{>}({\Bf k})}\,
\delta_{{\bF\Phi}_{\sigma}^{(>)}({\tilde{\bF k}}),{\bF k}},
&\;\; {\tilde{\Bf k}} \in \overline{\rm FS}_{\sigma}, \;
{\Bf k} \in \mbox{\rm 1BZ}.
\end{array} \right. \!\!\!\!\!\!\!\!\!\!\!\!\!\!\! \\
\end{eqnarray}
It is trivially verified that ${\sf\bar f}_{{\tilde{\bF k}};\sigma}
({\Bf k})$ reproduces the result in Eq.~(\ref{e15}) for $U\to 0$ and 
in addition yields (below ${\Bf k}, {\Bf k}' \in \mbox{\rm 1BZ}$)
\begin{eqnarray}
\label{e31}
&&\sum_{{\tilde{\bF k}}\in \mbox{\rm 1BZ} }
{\sf\bar f}_{{\tilde{\bF k}};\sigma}^*({\Bf k})
{\sf\bar f}_{{\tilde{\bF k}};\sigma}({\Bf k}')
= \delta_{{\bF k},{\bF k}'}, \\
\label{e32}
&&\sum_{{\tilde{\bF k}}\in 
{\rm FS}_{\sigma}/\,\ol{\rm FS}_{\sigma} }
{\sf\bar f}_{{\tilde{\bF k}};\sigma}^*({\Bf k})
{\sf\bar f}_{{\tilde{\bF k}};\sigma}({\Bf k}')
= \nu_{\sigma}^{\Ieq><}({\Bf k})\, \delta_{{\bF k},{\bF k}'}, 
\end{eqnarray}
in full conformity with Eqs.~(\ref{e13}) and (\ref{e16}) respectively. 
The significance of the mappings ${\Bf\Phi}_{\sigma}^{(\Ieq><)}
({\Bf k})$ becomes apparent by considering for a moment the possibility 
that ${\Bf\Phi}_{\sigma}^{(\Ieq><)}({\Bf k})$ are the identity mappings 
from the 1BZ on to itself. In this case, 
$\delta_{{\bF\Phi}_{\sigma}^{(\Ieq><)}({\tilde{\bF k}}), {\bF k}}$ on 
the right-hand side (RHS) of Eq.~(\ref{e30}) would be 
$\delta_{{\tilde{\bF k}},{\bF k}}$ and, as a consequence, Eq.~(\ref{e16}) 
corresponding to $\sum_s^{<}$ ($\sum_s^{>}$) would be reproduced by 
Eq.~(\ref{e32}) corresponding to the summation over ${\tilde{\Bf k}} 
\in {\rm FS}_{\sigma}$ (${\tilde{\Bf k}} \in \ol{\rm FS}_{\sigma}$) 
provided ${\Bf k}, {\Bf k}' \in {\rm FS}_{\sigma}$ (${\Bf k}, {\Bf k}' 
\in \overline{\rm FS}_{\sigma}$); the RHS of Eq.~(\ref{e32}) would be 
identically vanishing otherwise. Although, for weakly-interacting 
systems, $\nu_{\sigma}^{<}({\Bf k})$ and $\nu_{\sigma}^{>}({\Bf k})$ 
are small for ${\Bf k} \in \overline{\rm FS}_{\sigma}$ and ${\Bf k} 
\in {\rm FS}_{\sigma}$ respectively, this is not the case in 
general, establishing the significance of 
${\Bf\Phi}_{\sigma}^{(\Ieq><)}({\Bf k})$ as introduced above. Note 
that insofar as the expressions in Eqs.~(\ref{e31}) and (\ref{e32}) 
are concerned ${\Bf\Phi}_{\sigma}^{(<)}({\Bf k})$ and 
${\Bf\Phi}_{\sigma}^{(>)}({\Bf k})$ need {\sl not} be any further 
specified than they be one to one over the regions indicated in 
Eqs.~(\ref{e26}) and (\ref{e27}) respectively. 

\subsection*{\S~6. \sc A fictitious Green function and reproduction 
of some exact results}

Now we define the following fictitious single-particle Green 
function $\wt{\cal G}_{\sigma}({\Bf k};z)$, associated with the 
suppression of the `parameter of degeneracy' as indicated in 
Eq.~(\ref{e25}):
\begin{equation}
\label{e33}
\wt{\cal G}_{\sigma}({\Bf k};z)
{:=} \hbar \sum_{{\tilde{\bF k}} \in {\rm 1BZ}}
\frac{\vert {\sf\bar f}_{{\tilde{\bF k}};\sigma}({\Bf k})\vert^2 }
{z - \varepsilon_{{\tilde{\bF k}};\sigma} },
\end{equation}
which is to be compared with $\wt{G}_{\sigma}({\Bf k};z)$ in
Eq.~(\ref{e4}) above. Making use of the expressions in Eq.~(\ref{e30}),
we have
\begin{equation}
\label{e34}
\wt{\cal G}_{\sigma}({\Bf k};z) = \hbar \Big\{
\frac{ {\sf n}_{\sigma}({\Bf k})}{z -
\varepsilon_{{\bF\Phi}_{\sigma}^{(<) -1}({\bF k});\sigma} } +
\frac{ 1 - {\sf n}_{\sigma}({\Bf k}) }{ z -
\varepsilon_{{\bF\Phi}_{\sigma}^{(>) -1}({\bF k});\sigma} }
\Big\}.
\end{equation}
Here ${\Bf\Phi}_{\sigma}^{(\Ieq><) -1}({\Bf k})$ stand for the
inverse of ${\Bf\Phi}_{\sigma}^{(\Ieq><)}({\Bf k})$; they exist by 
our assumption with regard to the one-to-one property of the latter 
mappings. Equation (\ref{e34}) resembles the equation 
for the Green function in the atomic limit \cite{JHI63} in which, 
however, $n_{\sigma} = N_{\sigma}/N_{\sc l}$ is replaced by 
${\sf n}_{\sigma}({\Bf k})$. By analogy with the exact case, we 
define ({\it cf}. Eq.~(\ref{e18}) above)
\begin{equation}
\label{e35}
{\cal FS}_{\sigma} {:=} \{ {\Bf k} \, \| \,
1/{\cal G}_{\sigma}({\Bf k};{\tilde\mu}) > 0 \},
\end{equation}
and similarly for $\ol{\cal FS}_{\sigma}$. We specify $\tilde\mu$
in Eq.~(\ref{e42}) below.

We now demonstrate that ${\cal G}_{\sigma}({\Bf k};\varepsilon)$ 
yields the characteristics (i) - (iv) enumerated above. To this 
end, it is necessary first to determine the microscopic origin of 
$\{ {\sf\bar f}_{{\tilde{\bF k}};\sigma}({\Bf k}) \}$ and 
$\{ \varepsilon_{{\bF\Phi}_{\sigma}^{(\Ieq<>) -1}({\bF k});\sigma}\}$. 
In accordance with the replacement $s \rightharpoonup {\tilde{\Bf k}}$ 
in Eq.~(\ref{e25}) above, we denote the corresponding 
$\vert\Psi_{N_{\sigma} \pm 1,N_{\bar\sigma};s}\rangle$ by 
$\vert\Psi_{N_{\sigma}\pm 1,N_{\bar\sigma};{\tilde{\bF k}}}\rangle$. 
It is readily verified that, by defining 
$\vert\Psi_{N_{\sigma}\pm 1, N_{\bar\sigma};{\tilde{\bF k}}}\rangle$ 
according to
\begin{eqnarray}
\label{e36}
\vert\Psi_{N_{\sigma}-1,N_{\bar\sigma};{\tilde{\bF k}}} \rangle
{:=} \frac{1}{\sqrt{\nu_{\sigma}^{<}\big({\Bf\Phi}_{\sigma}^{(<)}
({\tilde{\Bf k}})\big)}}\,
&&{\hat c}_{{\bF\Phi}_{\sigma}^{(<)}
({\tilde{\bF k}});\sigma}\,
\vert\Psi_{N;0}\rangle,\; \nonumber\\
&&\;\;\;\;\;
{\tilde{\Bf k}} \in \mbox{\rm FS}_{\sigma},
\nonumber\\
\vert\Psi_{N_{\sigma}+1,N_{\bar\sigma};{\tilde{\bF k}}} \rangle
{:=} \frac{1}{\sqrt{\nu_{\sigma}^{>}\big({\Bf\Phi}_{\sigma}^{(>)}
({\tilde{\Bf k}})\big)}}\,
&&{\hat c}^{\dag}_{{\bF\Phi}_{\sigma}^{(>)}
({\tilde{\bF k}});\sigma}\,
\vert\Psi_{N;0}\rangle, \; \nonumber\\
&&\;\;\;\;\;
{\tilde{\Bf k}} \in \overline{\mbox{\rm FS}}_{\sigma},
\end{eqnarray}
${\sf\bar f}_{{\tilde{\bF k}};\sigma}({\Bf k})$ in Eq.~(\ref{e30})
directly follows from the defining expression in Eq.~(\ref{e5}) in 
which the `parameter of degeneracy' $\alpha$ in $s=(\tilde{\Bf k},
\alpha)$ is suppressed, in accordance with the substitutions in 
Eq.~(\ref{e25}). It is further readily verified that
\begin{equation} 
\label{e37}
\langle\Psi_{N_{\sigma}\pm 1,N_{\bar\sigma};{\tilde{\bF k}}}
\vert\Psi_{N_{\sigma}\pm 1,N_{\bar\sigma};{\tilde{\bF k}}'} \rangle
= \delta_{{\tilde{\bF k}},{\tilde{\bF k}}'},
\end{equation}
in conformity with $\langle\Psi_{M;s} \vert\Psi_{M,s'}\rangle = 
\delta_{s,s'}$. In arriving at the result in Eq.~(\ref{e37}) we 
have made use of the fact that ${\Bf\Phi}_{\sigma}^{(\Ieq<>)}
({\tilde{\Bf k}})$ are one to one so that 
$\delta_{{\bF\Phi}_{\sigma}^{(\Ieq<>)}({\tilde{\bF k}}), 
{\bF\Phi}_{\sigma}^{(\Ieq<>)}({\tilde{\bF k}}')} \equiv 
\delta_{{\tilde{\bF k}},{\tilde{\bF k}}'}$. 
Equations (\ref{e36}) concerning the $(N_{\sigma} \pm 1 + 
N_{\bar\sigma})$-particle `eigenstates' of $\wh{\cal H}$ give rise 
to the following `eigen-energies':
\footnote{\label{f2}
It can be explicitly shown \protect\cite{BF02} that in general the 
states in Eq.~(\protect\ref{e36}) are {\sl not} eigenstates of 
$\wh{\cal H}$ and thus ${\cal E}_{N_{\sigma}\pm 1,N_{\bar\sigma};
{\tilde{\bF k}}}$ are not in general eigenenergies of $\wh{\cal H}$. }
\begin{equation}
\label{e38}
{\cal E}_{N_{\sigma}\pm 1,N_{\bar\sigma};{\tilde{\bF k}}} {:=}
\langle\Psi_{N_{\sigma}\pm 1,N_{\bar\sigma};{\tilde{\bF k}}}
\vert \wh{\cal H}
\vert\Psi_{N_{\sigma}\pm 1,N_{\bar\sigma};{\tilde{\bF k}} }\rangle.
\end{equation}
Let
\footnote{\label{f3}
The subscript $0$ in ${\cal E}_{N_{\sigma}\pm 1,N_{\bar\sigma};0}$
is symbolic (analogous to $0$ in, for instance, $E_{N;0}$) and
does not denote $\tilde{\Bf k} = {\Bf 0}$. }
\begin{equation}
\label{e39}
{\cal E}_{N_{\sigma}\pm 1,N_{\bar\sigma};0} {:=}
\inf_{{\tilde{\bF k}} \in \ol{\rm FS}_{\sigma}
/\, {\rm FS}_{\sigma}} \big\{
{\cal E}_{N_{\sigma}\pm 1,N_{\bar\sigma};{\tilde{\bF k}}} \big\},
\end{equation}
for which by the variational principle we have
\begin{equation}
\label{e40}
E_{N_{\sigma}\pm 1,N_{\bar\sigma};0} \le
{\cal E}_{N_{\sigma}\pm 1,N_{\bar\sigma};0}.
\end{equation}
Further let ({\it cf}. Eq.~(\ref{e7}) and see Eq.~(\ref{e52}) below)
\begin{eqnarray}
\label{e41}
&&\tilde\mu_{N;\sigma}^{-} {:=} E_{N;0} - {\cal E}_{N_{\sigma}-1,
N_{\bar\sigma};0}, \nonumber \\
&&\tilde\mu_{N;\sigma}^{+} {:=} {\cal E}_{N_{\sigma}+1,
N_{\bar\sigma};0} - E_{N;0}.
\end{eqnarray}
The `chemical potential' $\tilde\mu$ introduced in Eq.~(\ref{e35})
has to satisfy ({\it cf}. Eq.~(\ref{e8}))
\begin{equation}
\label{e42}
\tilde\mu_{N;\sigma}^{-} < \tilde\mu <
\tilde\mu_{N;\sigma}^{+},\;\;\;
\forall \sigma.
\end{equation}
From Eqs.~(\ref{e7}), (\ref{e8}), (\ref{e40}) and (\ref{e41}),
one readily deduces that
\begin{equation}
\label{e43}
\tilde\mu_{N;\sigma}^{-} \le \mu_{N;\sigma}^{-} <
\mu_{N;\sigma}^{+} \le \tilde\mu_{N;\sigma}^{+}.
\end{equation}
It thus follows that, for both metallic and non-metallic GSs, $\tilde\mu$
in Eq.~(\ref{e35}) can be identified with $\mu$. It should be noted that, 
for cases where $\mu_{N;\sigma}^{+} - \mu_{N;\sigma}^{-}$ scales like
$1/N^p$, $p> 0$ (corresponding to metallic GSs, in macroscopic systems), 
Eq.~(\ref{e43}) in principle allows for $\tilde\mu_{N;\sigma}^{+} - 
\tilde\mu_{N;\sigma}^{-}$ to be finite, signifying a non-metallic GS 
associated with $\wt{\cal G}_{\sigma}({\Bf k};z)$. This is, however, 
not a viable possibility because, for {\sl macroscopic} metallic GSs, 
variational energies for the $(N_{\sigma}\pm 1 + 
N_{\bar\sigma})$-particle GSs deviate from their exact counterparts 
only by infinitesimal amounts (see also the considerations following
Eq.~(\ref{e60}) below). 

From the defining expression in Eq.~(\ref{e6}), following
Eq.~(\ref{e36}), for the fictitious single-particle excitation 
energies we have (for the poles of the fictitious Green function 
$\wt{\cal G}_{\sigma}({\Bf k};z)$ in Eq.~(\ref{e34}), see 
Eqs.~(\ref{e46}), (\ref{e47}) and (\ref{e48}) below)
\begin{eqnarray}
\label{e44}
\varepsilon_{{\tilde{\bF k}};\sigma} = \left\{ \begin{array}{ll}
E_{N;0} - {\cal E}_{N_{\sigma}-1,N_{\bar\sigma};{\tilde{\bF k}}},
&\;\;\; {\tilde{\Bf k}} \in {\rm FS}_{\sigma}, \\ \\
{\cal E}_{N_{\sigma}+1,N_{\bar\sigma};{\tilde{\bF k}}} - E_{N;0},
&\;\;\; {\tilde{\Bf k}} \in \overline{\rm FS}_{\sigma}.
\end{array} \right.
\end{eqnarray}
Note that, by construction $\varepsilon_{{\tilde{\bF k}};\sigma}
\le \tilde\mu_{N;\sigma}^{-} < \mu$ for ${\tilde{\Bf k}} \in
{\rm FS}_{\sigma}$ and $\mu < \tilde\mu_{N;\sigma}^{+} \le
\varepsilon_{{\tilde{\bF k}};\sigma}$ for ${\tilde{\Bf k}} \in
\overline{\rm FS}_{\sigma}$ (see Eqs.~(\ref{e39}) and (\ref{e41})). 
Thus, by construction (see also the considerations following 
Eq.~(\ref{e48}) below)
\begin{equation}
\label{e45}
{\cal FS}_{\sigma} \equiv {\rm FS}_{\sigma},\;\;\;
\overline{\cal FS}_{\sigma} \equiv \overline{\rm FS}_{\sigma}.
\end{equation}
From Eqs.~(\ref{e38}) and (\ref{e44}) we have 
\begin{equation}
\label{e46}
\varepsilon_{{\bF\Phi}_{\sigma}^{(\Ieq<>) -1}({\bF k});\sigma}
\equiv \varepsilon_{{\bF k};\sigma}^{\Ieq<>},\;\;\;
\forall {\Bf k} \in \mbox{\rm 1BZ},
\end{equation}
where 
\begin{eqnarray}
\label{e47}
&&\varepsilon_{{\bF k};\sigma}^{<}
= E_{N;0} - \frac{1}{\nu_{\sigma}^{<}({\Bf k})}\,
\langle\Psi_{N;0}\vert {\hat c}_{{\bF k};\sigma}^{\dag}
\wh{\cal H}\, {\hat c}_{{\bF k};\sigma} \vert\Psi_{N;0}\rangle,\\
\label{e48}
&&\varepsilon_{{\bF k};\sigma}^{>}
= \frac{1}{\nu_{\sigma}^{>}({\Bf k})}\,
\langle\Psi_{N;0}\vert {\hat c}_{{\bF k};\sigma}
\wh{\cal H}\, {\hat c}_{{\bF k};\sigma}^{\dag}
\vert\Psi_{N;0}\rangle - E_{N;0}.
\end{eqnarray}
The complete disappearance of any dependence on
${\Bf\Phi}^{(\Ieq><)}_{\sigma}({\Bf k})$ of these expressions, which 
is directly related to the assumed one-to-one nature of these mappings, 
is in full conformity with the results in Eq.~(\ref{e45}); on the
one hand, the definitions for ${\Bf\Phi}^{(\Ieq><)}_{\sigma}({\Bf k})$ 
in Eqs.~(\ref{e26}) and (\ref{e27}) explicitly require the knowledge 
of the exact FS$_{\sigma}$ while, on the other hand, the results 
in Eqs.~(\ref{e47}) and (\ref{e48}), which have {\sl no} {\sl explicit} 
dependence on ${\Bf\Phi}^{(\Ieq><)}_{\sigma}({\Bf k})$, establish this 
knowledge to be {\sl implicitly} present in the relevant functions
$\varepsilon_{{\bF k};\sigma}^{<}$ and $\varepsilon_{{\bF k};\sigma}^{>}$.

\subsection*{\S~7. \sc More exact results reproduced}

We are now in a position to expose the most significant properties
of ${\cal G}_{\sigma}({\Bf k};\varepsilon)$. To this end, let
${\cal A}_{\sigma}({\Bf k};\varepsilon)$ be the single-particle 
spectral function corresponding to ${\cal G}_{\sigma}({\Bf k};
\varepsilon)$ defined in Eq.~(\ref{e33}), according to
Eq.~(\ref{e9}). We have 
\begin{eqnarray}
\label{e49}
{\cal A}_{\sigma}({\Bf k};\varepsilon)
&=& \hbar \big\{
\nu_{\sigma}^{<}({\Bf k})\, \delta\big(\varepsilon -
\varepsilon_{{\bF k};\sigma}^{<} \big)
+ \nu_{\sigma}^{>}({\Bf k})\, \delta\big(\varepsilon -
\varepsilon_{{\bF k};\sigma}^{>} \big)
\big\},\nonumber\\
& &\;\;\;\;\;\;\;\;\;\;\;\;\;\;\;\;\;\;\;\;\;\;\;\;\;\;\;\;\;
\;\;\;\;\;\;\;\;\;\;\;\;\;\;
{\Bf k} \in \mbox{\rm 1BZ}.
\end{eqnarray}
Note in passing that in contrast with the non-interacting case, where 
$A_{\sigma}^{(0)}({\Bf k};\varepsilon) = \hbar \delta(\varepsilon 
- \epsilon_{{\bF k};\sigma})$, for {\sl all} ${\Bf k} \in 
{\rm 1BZ}$, ${\cal A}_{\sigma}({\Bf k};\varepsilon)$ consists of two 
$\delta$ functions, approaching, however, $A_{\sigma}^{(0)}({\Bf k};
\varepsilon)$ for $U \to 0$; since $\nu_{\sigma}^{<}({\Bf k}) + 
\nu_{\sigma}^{>} ({\Bf k}) = 1$ (see Eq.~(\ref{e17})) and 
$\varepsilon_{{\bF k};\sigma}^{<} < \mu < \varepsilon_{{\bF k};
\sigma}^{>}$, it follows that ${\cal A}_{\sigma}({\Bf k};\varepsilon)$ 
in a way mimics the behaviour of the exact $A_{\sigma}({\Bf k};
\varepsilon)$, according to which the particle-particle interaction 
gives rise to transfer of spectral weight from occupied to unoccupied 
states \cite{HL67}. From the above considerations, it is easily verified 
that ${\cal A}_{\sigma}({\Bf k};\varepsilon)$ also exactly reproduces 
the exact results in Eqs.~(\ref{e10}) and (\ref{e11}), as well as 
\begin{equation}
\label{e50}
\frac{1}{\hbar} \int_{-\infty}^{\infty} {\rm d}\varepsilon\;
\varepsilon\; A_{\sigma}({\Bf k};\varepsilon) =
\varepsilon_{\bF k} + U n_{\bar\sigma}.
\end{equation}
The fact that ${\cal A}_{\sigma}({\Bf k};\varepsilon)$ satisfies
the combination of Eqs.~(\ref{e10}) and (\ref{e11}) (i.e. that the 
$\varepsilon$ integral over $(-\infty,\infty)$ of $\hbar^{-1} 
{\cal A}_{\sigma}({\Bf k};\varepsilon)$ is equal to unity) as well as 
of Eq.~(\ref{e50}) implies that the first two leading-order terms in the 
large-$\vert\varepsilon\vert$ asymptotic series of ${\cal G}_{\sigma}
({\Bf k};\varepsilon)$ are {\sl identical} with those of $G_{\sigma}
({\Bf k};\varepsilon)$ \cite{BF01}. Further, for the GS total energy 
$E_{N;0}$ we have
\begin{equation}
\label{e51}
E_{N;0} = \frac{1}{2} \sum_{{\bF k},\sigma}
\varepsilon_{\bF k}\, {\sf n}_{\sigma}({\Bf k})
+ \frac{1}{2\hbar} \int_{-\infty}^{\mu}
{\rm d}\varepsilon\; \varepsilon\,
\sum_{{\bF k},\sigma} A_{\sigma}({\Bf k};\varepsilon),
\end{equation}
from which it follows that ${\cal E}_{N;0}$, the energy corresponding 
to the GS associated with ${\cal G}_{\sigma}({\Bf k};\varepsilon)$, 
is obtained through replacing $A_{\sigma}({\Bf k};\varepsilon)$ on 
the RHS of Eq.~(\ref{e51}) by ${\cal A}_{\sigma}({\Bf k};\varepsilon)$. 
By some straightforward algebra one can then deduce that 
\begin{equation}
\label{e52}
{\cal E}_{N;0} = E_{N;0}, 
\end{equation}
which shows an interesting aspect concerning the self-consistency 
of our scheme which in part relies on the definition for 
$\varepsilon_{{\tilde{\bF k}};\sigma}$ in Eq.~(\ref{e44}), through 
which it depends on $E_{N;0}$. 

From the defining expressions in Eqs.~(\ref{e47}) and (\ref{e48}), making 
use of the canonical anticommutation relations for ${\hat c}_{{\bF k};
\sigma}^{\dag}$ and ${\hat c}_{{\bF k};\sigma}$, we obtain
\begin{equation}
\label{e53}
\varepsilon_{{\bF k};\sigma}^{\Ieq<>}
= \varepsilon_{\bF k} 
+ U\, \frac{ \beta_{{\bF k};\sigma}^{\Ieq<>} }
{\nu_{\sigma}^{\Ieq<>}({\Bf k}) },\;\;\;
\forall {\Bf k} \in \mbox{\rm 1BZ},
\end{equation} 
where
\begin{eqnarray}
\label{e54}
&&\beta_{{\bF k};\sigma}^{<} \equiv
\frac{1}{N_{\sc l}}\,
\sum_{\sigma'} \sum_{{\bF p}',{\bF q}'} \nonumber\\ 
&&\;\;\;\;\;\;
\times\langle\Psi_{N;0}\vert {\hat c}_{{\bF k};\sigma}^{\dag}
{\hat c}_{{\bF p}'-{\bF q}';\sigma'}^{\dag}
{\hat c}_{{\bF p}';\sigma'}
{\hat c}_{{\bF k}-{\bF q}';\sigma}
\vert\Psi_{N;0}\rangle, \\ \nonumber \\
\label{e55}
&&\beta_{{\bF k};\sigma}^{>} \equiv n_{\bar\sigma} - 
\beta_{{\bF k};\sigma}^{<}.
\end{eqnarray}
The result in Eq.~(\ref{e55}) reflects the following {\sl exact}
relationship
\begin{equation}
\label{e56}
{\sf n}_{\sigma}({\Bf k})\, \varepsilon_{{\bF k};\sigma}^{<}
+ \big( 1 - {\sf n}_{\sigma}({\Bf k})\big) \,
\varepsilon_{{\bF k};\sigma}^{>} = \varepsilon_{\bF k} +
U\, n_{\bar\sigma},
\end{equation}
which is deduced through employing Eqs.~(\ref{e17}) and (\ref{e53}). 
On replacing $\vert\Psi_{N;0}\rangle$ in Eq.~(\ref{e54}) by an 
uncorrelated $N$-particle state, 
$\vert\Phi_{N;0}\rangle$, explicit calculation reveals that 
\begin{equation}
\label{e57}
\left.\varepsilon_{{\bF k};\sigma}^{\Ieq><}
\right|_{\vert\Phi_{N;0}\rangle }
= \varepsilon_{\bF k} + U\, n_{\bar\sigma}\;\;\;
\mbox{\rm for}\;\;\;
{\Bf k} \in {\rm FS}_{\sigma}^{(0)}, 
\ol{\rm FS}_{\sigma}^{(0)}, 
\end{equation}
where $n_{\bar\sigma}$ pertains to $\vert\Phi_{N;0}\rangle$. Note 
that it is due to the fundamental difference in the mappings defined 
in Eqs.~(\ref{e26}), (\ref{e27}) and Eqs.~(\ref{e28}), (\ref{e29}) that, 
in the `non-interacting' case, $\varepsilon_{{\bF k};\sigma}^{<}$ is 
defined over FS$_{\sigma}^{(0)}$ (as opposed to the entire 1BZ), and 
$\varepsilon_{{\bF k};\sigma}^{>}$ over $\ol{\rm FS}_{\sigma}^{(0)}$. 

Having established some fundamental (and, for our present considerations,
relevant) aspects of ${\cal G}_{\sigma}({\Bf k};\varepsilon)$, we are 
now capable of exposing a number of exact properties of the metallic 
GSs of the Hubbard Hamiltonian. In line with our above considerations, 
we restrict ourselves to 
GSs whose spatial periodicity is that implied by the underlying 
lattice $\{ {\Bf R}_j \}$. With
\begin{equation}
\label{e58}
\Lambda_{\sigma}({\Bf k}) {:=} 
\frac{ {\sf n}_{\sigma}({\Bf k}) }
{1 - {\sf n}_{\sigma}({\Bf k}) },\;\;\;\;
\Gamma_{\sigma}({\Bf k}) {:=}
\frac{ \mu - \varepsilon_{{\bF k};\sigma}^{<} }
{\varepsilon_{{\bF k};\sigma}^{>} - \mu },
\end{equation}
from Eqs.~(\ref{e35}) and (\ref{e45}), making use of 
Eqs.~(\ref{e34}) and (\ref{e46}), we have
\begin{equation}
\label{e59}
{\rm FS}_{\sigma} = \{ {\Bf k} \, \| \,
\Lambda_{\sigma}({\Bf k}) > \Gamma_{\sigma}({\Bf k}) \}.
\end{equation}
Similar expressions, in terms of $\Lambda_{\sigma}({\Bf k})$ and 
$\Gamma_{\sigma}({\Bf k})$, are obtained for $\ol{\rm FS}_{\sigma}$
and ${\cal S}_{{\sc f};\sigma}$. Following some algebra, for 
${\cal S}_{{\sc f};\sigma}$ we thus obtain
\begin{equation}
\label{e60}
{\cal S}_{{\sc f};\sigma} = 
\{ {\Bf k} \, \| \, (\varepsilon_{\sc f} - 
\varepsilon_{{\bF k};\sigma}^{<})
(\varepsilon_{\sc f}^+ - 
\varepsilon_{{\bF k};\sigma}^{>}) = 0 \},
\end{equation}
where $\varepsilon_{\sc f}^+ \equiv \mu_{N;\sigma}^+$. This together 
with the result in Eq.~(\ref{e42}), implying $\varepsilon_{{\bF k};
\sigma}^< < \varepsilon_{{\bF k};\sigma}^>$, allow for the following 
possibility when ${\Bf k}_{{\sc f};\sigma} \in {\cal S}_{{\sc f};\sigma}$ 
(recall that, according to Eq.~(\ref{e24}), ${\cal S}_{{\sc f};\sigma}$ 
{\sl may} be a proper subset of ${\cal S}_{{\sc f};\sigma}^{(0)}$): on 
the one hand $\varepsilon_{{\bF k}_{{\sc f};\sigma}^-;\sigma}^{<} =
\varepsilon_{\sc f}$, $\varepsilon_{{\bF k}_{{\sc f};\sigma}^-;\sigma}^{>}
= \varepsilon_{\sc f} + \Delta^{<}$ with $\Delta^{<} > 0$ and on the 
other $\varepsilon_{{\bF k}_{{\sc f};\sigma}^+;\sigma}^{>} =
\varepsilon_{\sc f}^+$, $\varepsilon_{{\bF k}_{{\sc f};\sigma}^+;\sigma}^{<} 
= \varepsilon_{\sc f} - \Delta^{>}$ with $\Delta^{>} > 0$. It is readily 
verified that these conditions in combination with the result in 
Eq.~(\ref{e56}) imply $(1 - {\sf n}_{\sigma}({\Bf k}_{{\sc f};\sigma}^-) 
\big)\, \Delta^{<} + {\sf n}_{\sigma}({\Bf k}_{{\sc f};\sigma}^+)\,
\Delta^{>} = 0$, which, unless $\Delta^{\Ieq><} = 0$ {\it or}
${\sf n}_{\sigma}({\Bf k}_{{\sc f};\sigma}^-)=1$ and ${\sf n}_{\sigma}
({\Bf k}_{{\sc f};\sigma}^+)=0$, cannot be satisfied. Thus for 
interacting metallic GSs, the two energy dispersions 
$\varepsilon_{{\bF k};\sigma}^{<}$ and $\varepsilon_{{\bF k};\sigma}^{>}$ 
coincide (up to infinitesimal corrections) with $\varepsilon_{\sc f}$ 
for ${\Bf k} \in {\cal S}_{{\sc f};\sigma}$. In other words, 
${\cal S}_{{\sc f};\sigma}$ in Eq.~(\ref{e60}) can be defined by either 
of the two conditions $\varepsilon_{\sc f} - \varepsilon_{{\bF k};
\sigma}^{<} = 0$ and $\varepsilon_{\sc f}^+ - \varepsilon_{{\bF k};
\sigma}^{>} = 0$. From this result it follows that, for metallic 
GSs, the infimum of ${\cal E}_{N_{\sigma}\pm 1,N_{\bar\sigma};
{\tilde{\bF k}}}$ as defined in Eq.~(\ref{e38}) is achieved (not 
necessarily exclusively) for all ${\tilde{\Bf k}} \in 
{\cal S}_{{\sc f};\sigma}$.

The above observations in conjunction with Eq.~(\ref{e56}) lead 
us to the following significant result for the uniform metallic 
GSs of $\wh{\cal H}$:
\begin{equation}
\label{e61}
\varepsilon_{\sc f} = 
\varepsilon_{{\bF k}={\bF k}_{{\sc f};\sigma}}
+ U\, n_{\bar\sigma},\;\;\;\;
\forall\, {\Bf k}_{{\sc f};\sigma} \in {\cal S}_{{\sc f};\sigma}.
\end{equation}
Since $\varepsilon_{\sc f}$ is a constant, the anisotropy of
$\varepsilon_{\bF k}$ together with the constancy of $U\, 
n_{\bar\sigma}$ with respect to ${\Bf k}$ necessitate the fundamental 
general result presented in Eq.~(\ref{e24}). We point out that, 
on the basis of this result, for the case of half-filling (corresponding 
to $n_{\sigma}=n_{\bar\sigma} = 1/2$) where by convention (see text 
following Eq.~(\ref{e2})) we have $\varepsilon_{\bF k} = 0$ for 
${\Bf k} \in {\cal S}_{{\sc f};\sigma}^{(0)}$, from Eq.~(\ref{e61}) 
we deduce that
\begin{equation}
\label{e62}
\varepsilon_{\sc f} = \frac{1}{2}\, U
\;\;\;\;\; \mbox{\rm (at half filling)},
\end{equation}
which constitutes a well-known rigorous theorem for $d > 1$ 
\cite{MM68,LPM69}. The validity of the result in Eq.~(\ref{e62}) 
for $d=1$, with $n_{\sigma} = n_{\bar\sigma} = n/2$ infinitesimally 
different from $1/2$ (see \cite{HS72}), can be made plausible by 
employing Eqs.~(22) and (23) in the work by Lieb and Wu \cite{LW68}, 
which concern the exact solution of the Hubbard model for $d=1$ 
(specific to $\varepsilon_k = -2 t \cos(k)$, where $k$ is in units 
of the inverse lattice constant) at half-filling, from which one 
obtains that $(\mu_{N;\sigma}^+ + \mu_{N;\sigma}^-)/2 = U/2$ for 
{\sl all} $U$ and that, for $U\downarrow 0$, $\mu_{N;\sigma}^+ - 
\mu_{N;\sigma}^-$ approaches zero exponentially rapidly; for 
instance, for $U/t=1$ this gap is equal to $5.0 \times 10^{-3} t$, 
whereas for $U/t=0.1$ it is equal to $5.9 \times 10^{-6} t$.

\subsection*{\S~8. \sc A kinematic constraint concerning 
partially polarized uniform metallic states}

\begin{figure}[t!]
\protect
\centerline{
\psfig{figure=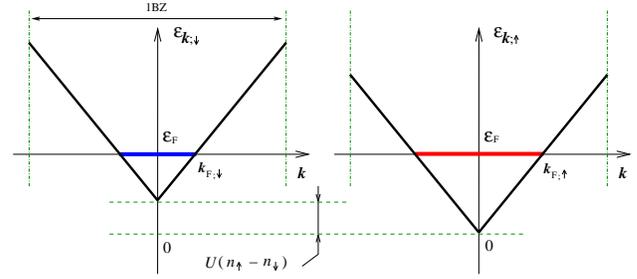,width=3.25in} }
\vskip 5pt
\caption{\label{fi2} \sf
Illustration of the kinematic constraint in 
Eq.~(\protect\ref{e63}). }
\end{figure}

The independence from $\sigma$ of $\varepsilon_{\sc f}$ together 
with the expression in Eq.~(\ref{e61}) imply the following result 
(Fig.~\ref{fi2}), applicable to uniform metallic GSs of $\wh{\cal H}$ 
corresponding to $N_{\sigma}, N_{\bar\sigma} \not= 0$:
\begin{equation}
\label{e63}
n_{\sigma} - n_{\bar\sigma} = \frac{1}{U}\,
\big( \varepsilon_{{\bF k}_{{\sc f};\sigma}} -
\varepsilon_{{\bF k}_{{\sc f};\bar\sigma}} \big)\;\;
\mbox{\rm when}\;\; n_{\sigma}, n_{\bar\sigma} \not=0;
\end{equation}
the conditions $n_{\sigma}, n_{\bar\sigma} \not= 0$ imply that 
Eq.~(\ref{e63}) has no bearing on fully 
ferromagnetic states, such as the Nagaoka state \cite{YN6566,DJT65}.
Leaving aside the fact that here $n_{\sigma}$ and $n_{\bar\sigma}$ 
(and in consequence, ${\Bf k}_{{\sc f};\sigma}$ and ${\Bf k}_{{\sc f};
\bar\sigma}$) correspond to the {\sl exact} $(N_{\sigma}+ 
N_{\bar\sigma})$-particle GS of $\wh{\cal H}$ (as opposed to an 
uncorrelated $(N_{\sigma}+N_{\bar\sigma})$-particle state), the 
condition in Eq.~(\ref{e63}) is equivalent to that within the 
framework of the Hartree-Fock scheme (see, e.g., Eq.~(7.6) in 
\cite{DRP66}). This aspect is immediately understood by considering
the fact that the Hartree-Fock formalism is a variational approach, 
in which the many-body Hamiltonian $\wh{\cal H}$ is exactly 
taken account of and the underlying approximation solely concerns 
the uncorrelated nature of the Hartree-Fock GS; in other words,
Eq.~(\ref{e63}) could have directly been inferred from its 
Hartree-Fock counterpart (i.e. from Eq.~(7.6) in \cite{DRP66}). 
To clarify this, let us consider Eq.~(\ref{e63}) within the 
framework of the Hartree-Fock scheme whereby its validity is 
undisputed. The appearance of $U$ on the RHS of Eq.~(\ref{e63}) has 
its origin in $\wh{\cal H}$, which, as we have mentioned above, 
is treated exactly within the Hartree-Fock framework. Consequently, 
accounting for the neglected correlation in the latter framework, 
which is {\sl implicit} in the exact GS $\vert\Psi_{N;0}\rangle$, 
{\sl cannot} entail introduction of contributions to
Eq.~(\ref{e63}) with {\sl explicit} dependence on $U$. It follows 
that the difference between the latter expression and its exact 
counterpart must reside in $n_{\sigma}$ (and consequently 
$n_{\bar\sigma}$ or equivalently, ${\Bf k}_{{\sc f};\sigma}$ and 
consequently ${\Bf k}_{{\sc f};\bar\sigma}$) which is an 
{\sl implicit} function of $U$. This demonstrates that the functional 
form of the expression in Eq.~(\ref{e63}) is universal and indeed 
Eq.~(\ref{e63}) is valid also within the framework of the exact theory. 
\footnote{\label{f4}
It appears that the presumed first-order transition from 
paramagnetic to fully ferromagnetic uniform GSs (and vice versa) 
within the Hartree-Fock framework must have been the cause for 
the widespread disregard of the counterpart of the constraint in 
Eq.~(\protect\ref{e63}) in Hartree-Fock calculations; 
Eq.~(\protect\ref{e63}) is always satisfied for $n_{\sigma}
=n_{\bar\sigma} = n/2$, and it does not apply to cases where
$n_{\sigma}=n$ and $n_{\bar\sigma}=0$, or $n_{\sigma}=0$ and 
$n_{\bar\sigma}=n$. Our careful numerical investigations \cite{BF02} 
reveal that there are instances where $n_{\sigma}\not=n_{\bar\sigma}$, 
satisfying $n_{\sigma}+n_{\bar\sigma}=n$, solve Eq.~(\protect\ref{e63})
and moreover yield the lowest variational energy in comparison with 
the energies of both paramagnetic and fully-ferromagnetic uniform 
states. This is specifically the case for instances where 
$\varepsilon_{\bF k}$ is expressed in terms of $t$ and $t'$ and for 
densities $n$ at and close to the associated van Hove densities. }

As we have indicated above, we relegate an extensive discussion
of the expression in Eq.~(\ref{e63}) to a future publication, in 
which we contrast the scarcity of the solutions $n_{\sigma} \not= 
n_{\bar\sigma}$ (satisfying $n_{\sigma} + n_{\bar\sigma} = n$) 
to Eq.~(\ref{e63}) with a wide range of results that clearly indicate 
a similar scarcity of ferromagnetic GSs for the single-band Hubbard 
Hamiltonian in particular for hyper-cubic lattices and strictly 
tight-binding $\varepsilon_{\bF k}$ for various $d$. For completeness,
for $d=1$, for the tight-binding energy dispersion $\varepsilon_k = 
-2 t\cos(k)$ (where $k$ is in units of the inverse lattice constant) 
Eq.~(\ref{e63}) takes on the form 
\begin{eqnarray}
n_{\sigma} - n_{\bar\sigma} = -\frac{2t}{U} 
\big[\cos(\pi n_{\sigma})-\cos(\pi n_{\bar\sigma})\big], 
\nonumber
\end{eqnarray}
which has to be solved in conjunction with the requirement $n_{\sigma}
+n_{\bar\sigma} = n$. Our analytic and numerical work on the latter 
transcendental equation \cite{BF02} reveals that this equation 
almost entirely accounts for the fact embodied by a theorem 
due to Lieb and Mattis \cite{LM62} according to which the total spin 
${\sf S}$ of the GS of $\wh{\cal H}$ for $d=1$ corresponding to 
$\varepsilon_k = -2 t\cos(k)$ and $U\not=\infty$ is equal to zero. 
Our numerical treatment of Eq.~(\ref{e63}) for $d=2$ corresponding to 
\begin{eqnarray}
\varepsilon_{\bF k} = -2 t \big[ \cos(k_x) 
+\cos(k_y) \big] + 4 t' \cos(k_x) \cos(k_y),
\nonumber
\end{eqnarray}
$t>0$, $t' \ge 0$, at and close to the van Hove density 
$n_{\rm v\sc h}$ (defined as the density for which the Fermi energy 
of the unpolarized state of the non-interacting fermions coincides 
with $\varepsilon_{\bF k}$ at ${\Bf k} = (\pi,0)$, or any of the 
symmetry-related ${\Bf k}$ points) are in conformity with the 
findings by Honerkamp and Salmhofer \cite{HS01a,HS01b} based on 
so-called temperature-flow renormalization-group scheme for 
$t'/t \approx 0.3$; for $t'=0$ and general $n$, our results conform 
with the Monte Carlo results obtained by Hirsch \cite{JEH85} and the
findings by Rudin and Mattis \cite{RM85}.

\subsection*{\S~9. \sc The zero-temperature limit of
${\sf n}_{\sigma}({\Bf k})$ for ${\Bf k}\in
{\cal S}_{{\sc f};\sigma}$ }

For metallic GSs corresponding to a $\Sigma_{\sigma}({\Bf k};
\varepsilon)$ which for ${\Bf k} \in {\cal S}_{{\sc f};\sigma}$ 
is continuously differentiable with respect to $\varepsilon$ in a 
neighbourhood of $\varepsilon=\varepsilon_{\sc f}$ (see condition
(A) presented in \S~11.2.2 below), we have $0 < 
Z_{{\bF k}_{{\sc f};\sigma}} \le 1$, where
\begin{equation}
\label{e64}
Z_{{\bF k}_{{\sc f};\sigma}} {:=}
{\sf n}_{\sigma}({\Bf k}_{{\sc f};\sigma}^-) -
{\sf n}_{\sigma}({\Bf k}_{{\sc f};\sigma}^+),
\end{equation}
which according to a Migdal theorem \cite{ABM57}, amounts to 
the strength of the coherent contribution to $A_{\sigma}({\Bf k};
\varepsilon)$ at $({\Bf k},\varepsilon) = ({\Bf k}_{{\sc f};\sigma}^-,
\varepsilon_{\sc f})$ (compare with $A_{\sigma}({\Bf k};\mu) \equiv
0$, $\forall {\Bf k}$). In cases where $Z_{{\bF k}_{{\sc f};\sigma}} 
\not= 0$, $\Lambda_{\sigma}({\Bf k})$ in Eq.~(\ref{e58}) is similar
to ${\sf n}_{\sigma}({\Bf k})$ discontinuous for ${\Bf k} \in 
{\cal S}_{{\sc f};\sigma}$. In spite of these, both ${\sf n}_{\sigma}
({\Bf k})$ and $\Lambda_{\sigma}({\Bf k})$ are well-defined functions 
for ${\Bf k}_{{\sc f};\sigma} \in {\cal S}_{{\sc f};\sigma}$; we 
have namely
\begin{equation}
\label{e65}
{\sf n}_{\sigma}({\Bf k}_{{\sc f};\sigma}) 
{:=} \frac{1}{2} \big[ 
{\sf n}_{\sigma}({\Bf k}_{{\sc f};\sigma}^{-}) +
{\sf n}_{\sigma}({\Bf k}_{{\sc f};\sigma}^{+}) \big],
\end{equation}
and similarly for $\Lambda_{\sigma}({\Bf k}_{{\sc f};\sigma})$. 
Equation (\ref{e65}) reflects the fact that in the theory 
of Fourier integral transformation, principal-value integrals are 
defined in accordance with the Cauchy prescription. Owing to the 
intrinsic symmetry of the Fermi function, the thermal momentum 
distribution function approaches the value in Eq.~(\ref{e65}) for 
the temperature $T$ approaching zero.

\subsection*{\S~10. \sc Pseudogap revisited; detailed characterization}

Before considering the behaviour of ${\sf n}_{\sigma}({\Bf k})$ for 
${\Bf k}\to {\cal S}_{{\sc f};\sigma}$ we expose two significant 
aspects associated with the case where ${\cal S}_{{\sc f};\sigma} 
\subset {\cal S}_{{\sc f};\sigma}^{(0)}$ ({\it cf}. Eq.~(\ref{e24})). 
To this end, let ${\Bf k}_{0} \in {\cal S}_{{\sc f};
\sigma}^{(0)}\backslash {\cal S}_{{\sc f};\sigma}$ and let 
${\Bf k}_{0}^-$ (${\Bf k}_{0}^+$) denote a vector infinitesimally close 
to ${\Bf k}_{0}$, located inside FS$_{\sigma}$ ($\ol{\rm FS}_{\sigma}$). 
With $\varepsilon_{\bF k}$ a continuous function of ${\Bf k}$, on 
account of Eq.~(\ref{e61}) the RHS of Eq.~(\ref{e56}) is equal to 
$\varepsilon_{\sc f}$ for ${\Bf k} = {\Bf k}_0^{\pm}$. In consequence 
of this, with $Z_{{\bF k}_{0};\sigma}$ defined according to 
Eq.~(\ref{e64}), from Eq.~(\ref{e56}) we obtain
\begin{eqnarray}
\label{e66}
&&Z_{{\bF k}_{0};\sigma} 
\big( \varepsilon_{{\bF k}_{0}^-;\sigma}^{<} -
\varepsilon_{{\bF k}_{0}^-;\sigma}^{>} \big)
= {\sf n}_{\sigma}({\Bf k}_{0}^+)\,
\big( \varepsilon_{{\bF k}_{0}^+;\sigma}^{<} -
\varepsilon_{{\bF k}_{0}^-;\sigma}^{<} \big) \nonumber \\
&&\;\;\;\;\;\;\;\;\;\;\;\;\;\;\;\;\;\;\;\;
+ \big( 1 - {\sf n}_{\sigma}({\Bf k}_{0}^+) \big)
\big( \varepsilon_{{\bF k}_{0}^+;\sigma}^{>} -
\varepsilon_{{\bF k}_{0}^-;\sigma}^{>} \big).
\end{eqnarray} 
Since ${\Bf k}_{0} \in {\cal S}_{{\sc f};\sigma}^{(0)} \backslash 
{\cal S}_{{\sc f};\sigma}$, by definition 
$\varepsilon_{{\bF k}_{0}^-;\sigma}^{<} - \varepsilon_{{\bF k}_{0}^-;
\sigma}^{>}$ must be a finite (negative) constant. Thus, by 
assuming $\varepsilon_{{\bF k};\sigma}^{\Ieq<>}$ to be continuous 
functions of ${\Bf k}$ in a neighbourhood of ${\Bf k}_{0}$, we observe 
that the RHS of Eq.~(\ref{e66}) must be vanishing. It follows that 
\begin{equation}
\label{e67}
Z_{{\bF k}_{0};\sigma}=0,\;\;\;\forall {\Bf k}_0 \in
{\cal S}_{{\sc f};\sigma}^{(0)}\backslash
{\cal S}_{{\sc f};\sigma},
\end{equation} 
revealing a significant aspect of ${\cal S}_{{\sc f};\sigma}^{(0)}
\backslash {\cal S}_{{\sc f};\sigma}$ which we have identified as 
the pseudogap region of the putative Fermi surface of the 
metallic GS of $\wh{\cal H}$. Now, since by the Hellmann-Feynman 
theorem
\begin{equation}
\label{e68}
{\sf n}_{\sigma}({\Bf k})  = 
\frac{\delta E_{N;0}}{\delta \epsilon_{{\bF k};\sigma}},
\end{equation}
it follows that ${\sf n}_{\sigma}({\Bf k})$ must be singular ({\sl not} 
necessarily discontinuous) at all ${\Bf k} \in {\cal S}_{{\sc f};
\sigma}^{(0)}$ \cite{BF02}; this assertion is based on the observation 
that to any non-vanishing variation $\delta \epsilon_{{\bF k};
\sigma}$ for ${\Bf k} \in {\cal S}_{{\sc f};\sigma}^{(0)}$ (in the 
present context $\delta\epsilon_{{\bF k};\sigma}$ is bound by the 
requirement that $\delta\epsilon_{{\bF k};\sigma} = 
\delta\epsilon_{{\bF\omega}{\bF k};\sigma}$ for {\sl all} elements 
of the set $\{ {\Bf\omega}\}$ of the point-group operations of the 
underlying lattice group) corresponds a Fermi surface 
${\cal S}_{{\sc f};\sigma}^{(0)\prime}$ (whose interior contains the 
same number of ${\Bf k}$ points as ${\cal S}_{{\sc f};\sigma}^{(0)}$, 
namely $N_{\sigma}$) that is {\sl not} capable of being obtained 
from ${\cal S}_{{\sc f};\sigma}^{(0)}$ through a continuous 
deformation; in this connection note that the functional derivative 
in Eq.~(\ref{e68}) is constrained by the requirement that the GS of 
$\wh{\cal H} + \delta\wh{\cal H}$, with $\delta \wh{\cal H} {:=} 
\sum_{\sigma} \delta\epsilon_{{\bF k};\sigma}\, \sum_{\bF\omega} 
{\hat c}_{{\bF\omega}{\bF k};\sigma}^{\dag} 
{\hat c}_{{\bF\omega}{\bF k};\sigma}$, whose energy we denote by 
$E_{N;0} + \delta E_{N;0}$, lie in the $(N_{\sigma}
+N_{\bar\sigma})$-particle Hilbert space of $\wh{\cal H}$. Thus 
${\sf n}_{\sigma}({\Bf k})$ is singular at ${\Bf k} = {\Bf k}_0$, 
from which we conclude that in the present case Eq.~(\ref{e12}) must 
possess at least two solutions at $\varepsilon=\varepsilon_{{\bF k}_0;
\sigma}^{(i)}$, $i=1,2$, for ${\Bf k}={\Bf k}_0 \in 
{\cal S}_{{\sc f};\sigma}^{(0)}\backslash {\cal S}_{{\sc f};\sigma}$, 
satisfying (see the paragraph following 
Eq.~(\ref{e24})) 
\begin{equation}
\label{e69}
\varepsilon_{{\bF k}_0;\sigma}^{(1)} < \varepsilon_{\sc f},\;\;\;\;\;
\varepsilon_{{\bF k}_0;\sigma}^{(2)} > \mu_{N;\sigma}^+.
\end{equation}
In the event ${\rm Im}[\Sigma_{\sigma}({\Bf k}_0;\varepsilon)]
\equiv 0$ for $\varepsilon\in [\varepsilon_{{\bF k}_0;\sigma}^{(1)},
\varepsilon_{\sc f}]$, by the variational principle we have (see 
our discussions corresponding to Eq.~(\ref{e43}))
\begin{equation}
\label{e70}
\varepsilon_{{\bF k}_0;\sigma}^{<}\, \le \,
\varepsilon_{{\bF k}_0;\sigma}^{(1)}.
\end{equation}
Similarly, in the event ${\rm Im}[\Sigma_{\sigma}({\Bf k}_0;
\varepsilon)]\equiv 0$ for $\varepsilon\in [\mu_{N;\sigma}^+,
\varepsilon_{{\bF k}_0;\sigma}^{(2)}]$, by the variational principle 
we have $\varepsilon_{{\bF k}_0;\sigma}^{>}\, \ge \, 
\varepsilon_{{\bF k}_0;\sigma}^{(2)}$. For macroscopic systems it is most 
likely that the two energies in Eq.~(\ref{e70}) (as well as the latter 
two energies) are within a small difference equal. Note that, since 
by assumption ${\Bf k}_0 \in {\cal S}_{{\sc f};\sigma}^{(0)}\backslash 
{\cal S}_{{\sc f};\sigma}$, Eq.~(\ref{e12}) cannot be satisfied 
at $({\Bf k};\varepsilon)=({\Bf k}_0;\varepsilon_{\sc f})$. We point out 
that {\sl only} when ${\rm Im}[\Sigma_{\sigma}({\Bf k}_0;\varepsilon)] 
\equiv 0$ for $\varepsilon \in [\varepsilon_{{\bF k}_0;\sigma}^{(1)},
\varepsilon_{{\bF k}_0;\sigma}^{(2)}]$ does the single-particle 
spectrum possess a direct gap at ${\Bf k}={\Bf k}_0$, whose value 
$\varepsilon_{{\bF k}_0;\sigma}^{(2)}-\varepsilon_{{\bF k}_0;
\sigma}^{(1)}$ does not exceed $\varepsilon_{{\bF k}_0;\sigma}^{>}
-\varepsilon_{{\bF k}_0;\sigma}^{<}$; when ${\rm Im}[\Sigma_{\sigma}
({\Bf k}_0;\varepsilon)] \equiv 0$ only for $\varepsilon$ in one of the 
intervals $[\varepsilon_{{\bF k}_0;\sigma}^{(1)},\varepsilon_{\sc f}]$ 
and $[\varepsilon_{\sc f},\varepsilon_{{\bF k}_0;\sigma}^{(2)}]$ is the 
gap in question is an indirect gap, not exceeding $\varepsilon_{\sc f}
-\varepsilon_{{\bF k}_0;\sigma}^{<}$ and $\varepsilon_{{\bF k}_0;
\sigma}^{>}-\varepsilon_{\sc f}$ respectively. Otherwise, although 
there is {\sl no} true gap in the single-particle excitation spectrum 
at ${\Bf k}_0 \in {\cal S}_{{\sc f};\sigma}^{(0)} \backslash 
{\cal S}_{{\sc f};\sigma}$, nonetheless the single-particle spectral 
function $A_{\sigma}({\Bf k}_0;\varepsilon)$ is expected to be relatively 
considerable in the neighbourhoods of $\varepsilon_{{\bF k}_0;\sigma}^{<}$ 
and $\varepsilon_{{\bF k}_0;\sigma}^{>}$ and suppressed in the intervening 
interval, rendering the designation `pseudogap' in the single-particle 
excitation spectrum appropriately descriptive. Note that, although 
$\varepsilon_{{\bF k}_0;\sigma}^{(i)}$, $i=1,2$, satisfy Eq.~(\ref{e12}), 
nonetheless since in the case at hand $Z_{{\bF k}_{0};\sigma}=0$
(see Eq.~(\ref{e67})), the spectral function $A_{\sigma}
({\Bf k}_0;\varepsilon)$ does not possess quasiparticle 
peaks at $\varepsilon=\varepsilon_{{\bF k}_0;\sigma}^{(i)}$, $i=1,2$, 
but {\sl resonances}. Experimentally \cite{AGL96,MDL96,MRN98,AI99}, 
this feature is characteristic of the under-doped cuprates for which, 
in the normal state, in certain regions of the ${\Bf k}$ space that 
conventionally would be considered as constituting the Fermi surface, 
the expected coherent quasiparticle peak in the angle-resolved 
photo-emission spectrum is missing, and, further, a gap (experimentally 
identified as the interval between $\varepsilon_{\sc f}$ and the 
`leading edge' of the broad peak in the angle-resolved photo-emission 
spectrum) in this spectrum persists for $T$ greater than the
superconducting transition temperature $T_{\rm c}$.

\subsection*{\S~11. \sc The behaviour of ${\sf n}_{\sigma}({\Bf k})$
for ${\Bf k} \to {\cal S}_{{\sc f};\sigma}$ }

Now we proceed with considering the behaviour of ${\sf n}_{\sigma}
({\Bf k})$ for ${\Bf k}$ close to ${\cal S}_{{\sc f};\sigma}$.
Equation (\ref{e53}) suggests that $\beta_{{\bF k};\sigma}^{<}$ 
should be best described as follows:
\begin{equation}
\label{e71}
\beta_{{\bF k};\sigma}^{<} = {\sf n}_{\sigma}({\Bf k})\,
\xi_{{\bF k};\sigma}.
\end{equation}
This form is general and does not impose any restriction on 
the behaviour of $\beta_{{\bF k};\sigma}^{<}$ in regions where 
${\sf n}_{\sigma}({\Bf k}) \not= 0$. Owing to Eq.~(\ref{e60})
(see also the text following this equation) and the result in 
Eq.~(\ref{e61}), for metallic states we must have
\begin{equation}
\label{e72}
\xi_{{\bF k};\sigma} = n_{\bar\sigma} +
\zeta_{{\bF k};\sigma},\;\;\mbox{\rm where}\;\;
\zeta_{{\bF k};\sigma} \sim 0\;\;\mbox{\rm for}\;\;
{\Bf k}\to {\cal S}_{{\sc f};\sigma}.
\end{equation}
Employing the following series (below, ${\Bf k}_{{\sc f};\sigma}$
is the point at which ${\Bf k}$ meets ${\cal S}_{{\sc f};\sigma}$ 
in the limit ${\Bf k} \to {\cal S}_{{\sc f};\sigma}$)
\begin{eqnarray}
\label{e73}
&&\varepsilon_{\bF k} \sim
\varepsilon_{{\bF k}_{{\sc f};\sigma}}
+ {\Bf a}({\Bf k}_{{\sc f};\sigma}) \cdot
({\Bf k} - {\Bf k}_{{\sc f};\sigma})\;\;\mbox{\rm for}\;\;
{\Bf k} \to {\cal S}_{{\sc f};\sigma}, \nonumber\\
&&{\Bf a}({\Bf k}_{{\sc f};\sigma}) {:=}
\left. {\Bf\nabla}_{\bF k}\varepsilon_{\bF k}\right|_{{\bF k}
= {\bF k}_{{\sc f};\sigma}}
\equiv \hbar\, {\Bf v}_{{\sc f};\sigma}^{(0)},
\end{eqnarray}
from Eqs.~(\ref{e53}), (\ref{e71}) and (\ref{e72}) we obtain
\begin{eqnarray}
\label{e74}
\varepsilon_{{\bF k};\sigma}^{<}
\sim \mu +{\Bf a}({\Bf k}_{{\sc f};\sigma})\cdot
({\Bf k} - {\Bf k}_{{\sc f};\sigma}) &+& U\, \zeta_{{\bF k};\sigma}
\nonumber\\
\mbox{\rm for}\;\; {\Bf k} &\to& {\cal S}_{{\sc f};\sigma}.
\end{eqnarray}
From Eqs.~(\ref{e53}), (\ref{e55}), (\ref{e71}) and (\ref{e72}) we 
similarly obtain
\begin{eqnarray}
\label{e75}
\varepsilon_{{\bF k};\sigma}^{>} 
\sim \mu + {\Bf a}({\Bf k}_{{\sc f};\sigma})\cdot
({\Bf k} - {\Bf k}_{{\sc f};\sigma}) &-& U\,
\Lambda_{\sigma}({\Bf k})\, \zeta_{{\bF k};\sigma} \nonumber\\
\mbox{\rm for}\;\, {\Bf k} &\to& {\cal S}_{{\sc f};\sigma},
\end{eqnarray}
where $\Lambda_{\sigma}({\Bf k})$ is defined in Eq.~(\ref{e58}). 
Thus from Eqs.~(\ref{e58}), (\ref{e74}) and (\ref{e75}) we have
\begin{equation}
\label{e76}
\Gamma_{\sigma}({\Bf k}) \sim
\frac{ -{\Bf a}({\Bf k}_{{\sc f};\sigma})
\cdot ({\Bf k} - {\Bf k}_{{\sc f};\sigma}) -
U\,\zeta_{{\bF k};\sigma} }
{ {\Bf a}({\Bf k}_{{\sc f};\sigma})
\cdot ({\Bf k} - {\Bf k}_{{\sc f};\sigma}) -
U\,\Lambda_{\sigma}({\Bf k})\,\zeta_{{\bF k};\sigma} }, \;\;
{\Bf k} \to {\cal S}_{{\sc f};\sigma}.
\end{equation}

Let now ({\it cf}. Eq.~(\ref{e72}))
\begin{equation}
\label{e77}
\vert\zeta_{{\bF k};\sigma} \vert \sim
\vert B({\Bf k}_{{\sc f};\sigma})\vert\,
\| {\Bf k} - {\Bf k}_{{\sc f};\sigma} \|^{\gamma}\;\;
\mbox{\rm for}\;\;
{\Bf k} \to {\cal S}_{{\sc f};\sigma}.
\end{equation}
We consider three cases: case I, $0 < \gamma < 1$; case II,
$\gamma=1$; case III, $\gamma > 1$. The findings corresponding 
to case I equally apply to cases where, for instance,
\begin{equation}
\label{e78}
\vert\zeta_{{\bF k};\sigma} \vert \sim
\left| B({\Bf k}_{{\sc f};\sigma})
\ln\| {\Bf k} - {\Bf k}_{{\sc f};\sigma} \| \right|\,
\| {\Bf k} - {\Bf k}_{{\sc f};\sigma} \|^{\gamma}\;
\mbox{\rm for}\;
{\Bf k} \to {\cal S}_{{\sc f};\sigma}.
\end{equation}
In fact, the results corresponding to case I also apply to the 
case corresponding to $\gamma=1$ when, instead of Eq.~(\ref{e77}), 
Eq.~(\ref{e78}) holds; in such an event, our considerations 
corresponding to $\gamma=1$ become redundant. We emphasize that 
in principle the functional form of $\zeta_{{\bF k};\sigma}$ for 
${\Bf k}\to {\cal S}_{{\sc f};\sigma}$ can depend on the location 
of ${\Bf k}$, whether inside or outside ${\rm FS}_{\sigma}$. Thus, 
for instance, the behaviour of $\zeta_{{\bF k};\sigma}$ for 
${\Bf k}\to {\cal S}_{{\sc f};\sigma}$ may be governed by two 
powers $\gamma^{<}$ and $\gamma^{>}$, with $\gamma^{<} \not= 
\gamma^{>}$, for ${\Bf k}\in {\rm FS}_{\sigma}$ and ${\Bf k}\in 
\overline{\rm FS}_{\sigma}$ respectively. We shall not explicitly 
deal with such instances, however our following analyses can be 
readily extended to cover these. 

For completeness, we mention that following some algebra 
(involving for instance use of the equation of motion for 
${\hat c}_{{\bF k};\sigma}$ in the Heisenberg picture) we obtain
\begin{equation}
\label{e79}
\zeta_{{\bF k};\sigma} =
\frac{1}{U} \Big\{ 
\frac{\int_{-\infty}^{\mu}
{\rm d}\varepsilon\;\varepsilon\, A_{\sigma}({\Bf k};\varepsilon)}
{\int_{-\infty}^{\mu}
{\rm d}\varepsilon\; A_{\sigma}({\Bf k};\varepsilon)} 
-\varepsilon_{\bF k} \Big\} - n_{\bar\sigma}.
\end{equation}
Making use of Eqs.~(\ref{e61}), (\ref{e64}) and (\ref{e72}), from 
Eq.~(\ref{e79}) we readily deduce
\begin{equation}
\label{e80}
Z_{{\bF k}_{{\sc f};\sigma}} = \frac{1}{\hbar\varepsilon_{\sc f}}
\int_{-\infty}^{\mu} {\rm d}\varepsilon\;
\varepsilon \big[ 
A_{\sigma}({\Bf k}_{{\sc f};\sigma}^-;\varepsilon) -
A_{\sigma}({\Bf k}_{{\sc f};\sigma}^+;\varepsilon) \big],
\end{equation}
which can be of experimental significance (see, e.g., \cite{CGD99}) 
specifically in conjunction with the well-known result, directly 
deduced from Eqs.~(\ref{e10}) and (\ref{e64}), namely
\begin{equation}
\label{e81}
Z_{{\bF k}_{{\sc f};\sigma}} = \frac{1}{\hbar}
\int_{-\infty}^{\mu} {\rm d}\varepsilon\;
\big[ A_{\sigma}({\Bf k}_{{\sc f};\sigma}^-;\varepsilon) -
A_{\sigma}({\Bf k}_{{\sc f};\sigma}^+;\varepsilon) \big];
\end{equation}
although the experimentally 
determined $A_{\sigma}({\Bf k}_{{\sc f};\sigma}^{\mp};\varepsilon)$ 
may be too crude for the above expressions to provide reliable values 
for $Z_{{\bF k}_{{\sc f};\sigma}}$ (it is, however, natural to expect 
that, in view of the smallness of $\| {\Bf k}_{{\sc f};\sigma}^+ - 
{\Bf k}_{{\sc f};\sigma}^-\|$, the difference $A_{\sigma}
({\Bf k}_{{\sc f};\sigma}^-;\varepsilon) - A_{\sigma}
({\Bf k}_{{\sc f};\sigma}^+;\varepsilon)$ should be relatively 
accurate), nonetheless the knowledge that $0 \le 
Z_{{\bF k}_{{\sc f};\sigma}} < 1$ enables one to enforce, through 
use of Eqs.~(\ref{e80}) and (\ref{e81}), some degree of accuracy 
in the analyses of the measured $A_{\sigma}({\Bf k}_{{\sc f};
\sigma}^{\mp};\varepsilon)$. It is relevant to note that 
Eqs.~(\ref{e80}) and (\ref{e81}) remain valid by substituting 
$A_{\sigma}$ herein by ${\cal A}_{\sigma}$ (see Eq.~(\ref{e49})), 
this on account of the continuity of $\varepsilon_{{\bF k};
\sigma}^{\Ieq><}$ in a neighbourhood of ${\cal S}_{{\sc f};\sigma}$ 
and of $\varepsilon_{{\bF k};\sigma}^{<} =\varepsilon_{\sc f}$ for 
${\Bf k} \in {\cal S}_{{\sc f};\sigma}$ and $\varepsilon_{{\bF k};
\sigma}^{>} > \mu$ for ${\Bf k} \in {\rm 1BZ}$. 

\subsubsection*{{\rm 11.1.}~Case I: $\; 0 < \gamma < 1$}

In this case, from Eq.~(\ref{e76}) we immediately obtain
\begin{equation}
\label{e82}
\Gamma_{\sigma}({\Bf k}) \sim
\frac{1}{\Lambda_{\sigma}({\Bf k}) } \;\;\;
\mbox{\rm for}\;\;
{\Bf k} \to {\cal S}_{{\sc f};\sigma}.
\end{equation}
Solving for the Fermi surface ${\cal S}_{{\sc f};\sigma}$, that
is solving (see Eq.~(\ref{e59}) and the subsequent text)
\begin{equation}
\label{e83}
\Lambda_{\sigma}({\Bf k}) = \Gamma_{\sigma}({\Bf k}),
\end{equation}
from the result in Eq.~(\ref{e82}) we obtain (note that by definition 
$\Lambda_{\sigma}({\Bf k}) \ge 0$ so that the negative solution 
$\Lambda_{\sigma}({\Bf k}_{{\sc f};\sigma})=-1$ has to be discarded)
\begin{equation}
\label{e84}
\Lambda_{\sigma}({\Bf k}_{{\sc f};\sigma}) = 1,
\end{equation}
which through the defining expression for $\Lambda_{\sigma}({\Bf k})$ 
in Eq.~(\ref{e58}), according to which
\begin{equation}
\label{e85}
{\sf n}_{\sigma}({\Bf k}) 
= \frac{\Lambda_{\sigma}({\Bf k})}
{1 + \Lambda_{\sigma}({\Bf k})},
\end{equation}
results in
\begin{equation}
\label{e86}
{\sf n}_{\sigma}({\Bf k}_{{\sc f};\sigma}) = \frac{1}{2}.
\end{equation}
Note the significant fact that in the case under consideration,
${\sf n}_{\sigma}({\Bf k})$ is continuous in neighbourhood of 
${\Bf k}={\Bf k}_{{\sc f};\sigma}$, so that (see Eq.~(\ref{e64}) and 
our remark with regard to the significance of the following result)
\begin{equation}
\label{e87}
Z_{{\bF k}_{{\sc f};\sigma}}=0. 
\end{equation}
We point out that for the Hubbard (at half-filling 
\footnote{\label{f4a}
Concerning the continuity at $k=\mp k_{{\sc f};\sigma}$
of ${\sf n}_{\sigma}(k)$ for $d=1$, in the half-filled case,
we refer the reader to \protect\cite{BF03}. }
or $U/t \to \infty$) (see \cite{OS90,MT77,JV93,JS79}) and the 
Tomonaga model (see \cite{GS68}) for $d=1$, both 
${\sf n}_{\sigma}(\pm k_{{\sc f};\sigma})=1/2$ and 
$Z_{\pm k_{{\sc f};\sigma}}=0$. 

The result in Eq.~(\ref{e87}) implies that in the case under 
consideration, $\Sigma_{\sigma}({\Bf k};\varepsilon)$, with 
${\Bf k}\in {\cal S}_{{\sc f};\sigma}$, is not a continuously 
differentiable function of $\varepsilon$ in a neighbourhood of 
$\varepsilon=\varepsilon_{\sc f}$ \cite{BF99a}. As an explicit 
calculation shows \cite{BF99a} (see appendix D herein), for the 
one-dimensional Luttinger model \cite{JML63,ML65} we in addition 
have that ${\rm d}\wt{\Sigma}_{\sigma}(k;z)/{\rm d}k$ diverges for 
{\sl all} $z$ as $k \to \mp k_{{\sc f};\sigma}$. Equations 
(\ref{e77}) and (\ref{e78}), with $0 < \gamma < 1$ and 
$0 < \gamma \le 1$ respectively, are suggestive of a similar 
singular behaviour in ${\Bf\nabla}_{\bF k} \wt{\Sigma}_{\sigma}
({\Bf k};z)$ as ${\Bf k} \to {\cal S}_{{\sc f};\sigma}$. We 
therefore conclude that the identification with ``Luttinger-liquid''
by Anderson \cite{PWA90a,PWA90b,PWA97} of the metallic states of 
the Hubbard Hamiltonian for (specifically) $d=2$, entails that the 
case considered here (concerning $0 < \gamma < 1$, or, in the event 
Eq.~(\ref{e78}) applies, $0 < \gamma \le 1$) should encompass all 
uniform metallic GSs of the Hubbard Hamiltonian.

\subsubsection*{{\rm 11.2.}~Case II: $\; \gamma = 1$}
\noindent{11.2.1. \it General}
\vspace{0.2cm}

For the considerations in this case, it is crucial to be explicit 
with regard to the location of ${\Bf k}$ (i.e. whether ${\Bf k} \in 
{\rm FS}_{\sigma}$ or ${\Bf k} \in \overline{\rm FS}_{\sigma}$) for 
${\Bf k} \to {\cal S}_{{\sc f};\sigma}$. This is established 
by determining the sign of the inner product $\hat{\Bf n}
({\Bf k}_{{\sc f};\sigma}) \cdot ({\Bf k} - {\Bf k}_{{\sc f};\sigma})$,
where $\hat{\Bf n}({\Bf k}_{{\sc f};\sigma})$ stands for the outward
unit vector normal to ${\cal S}_{{\sc f};\sigma}$ at ${\Bf k} = 
{\Bf k}_{{\sc f};\sigma}$, pointing from FS$_{\sigma}$ to 
$\overline{\rm FS}_{\sigma}$. In order to simplify our subsequent 
analyses, unless we explicitly indicate otherwise, in what follows we 
assume ${\Bf k}-{\Bf k}_{{\sc f};\sigma}$ to be collinear with 
$\hat{\Bf n}({\Bf k}_{{\sc f};\sigma})$; that is, for sufficiently 
small $\| {\Bf k} - {\Bf k}_{{\sc f};\sigma}\|$ we assume that
\begin{equation}
\label{e88}
{\Bf k}-{\Bf k}_{{\sc f};\sigma} =
\mp \| {\Bf k}-{\Bf k}_{{\sc f};\sigma} \| \,
\hat{\Bf n}({\Bf k}_{{\sc f};\sigma}),\;\;
{\Bf k} \in \mbox{\rm FS}_{\sigma},\, 
\overline{\mbox{\rm FS}}_{\sigma}.
\end{equation}
Accordingly, we introduce
\begin{equation}
\label{e89}
{\Bf k}_{{\sc f};\sigma}^{\pm} {:=}
{\Bf k}_{{\sc f};\sigma} \pm \kappa\, 
\hat{\Bf n}({\Bf k}_{{\sc f};\sigma}),
\;\;\; \kappa\downarrow 0.
\end{equation}

Following Eq.~(\ref{e72}), in the case at hand for ${\Bf k}\to
{\Bf k}_{{\sc f};\sigma}$ we have ({\it cf}. Eq.~(\ref{e77}))
\begin{equation}
\label{e90}
\zeta_{{\bF k};\sigma} \sim
{\Bf b}_{\sigma}({\Bf k}_{{\sc f};\sigma}^{\mp}) \cdot
({\Bf k}-{\Bf k}_{{\sc f};\sigma})\;\;\;
\mbox{\rm for}\;\;
{\Bf k} \in \mbox{\rm FS}_{\sigma},\, 
\overline{\mbox{\rm FS}}_{\sigma}.
\end{equation}
The vector ${\Bf b}_{\sigma}({\Bf k})$ is {\sl not} a smoothly-varying 
function of ${\Bf k}$ as ${\Bf k}$ is transposed from inside 
FS$_{\sigma}$ through ${\cal S}_{{\sc f};\sigma}$ into 
$\overline{\rm FS}_{\sigma}$, and hence our use of ${\Bf k}_{{\sc f};
\sigma}^{\mp}$ as the argument of the vector function in Eq.~(\ref{e90}). 
For conciseness, in what follows we employ the notation
\begin{equation}
\label{e91}
{\Bf b}_{\sigma}^{\mp} {:=} 
{\Bf b}_{\sigma}({\Bf k}_{{\sc f};\sigma}^{\mp}).
\end{equation}
We point out that the possibility of $\|{\Bf b}_{\sigma}^{\mp}\|=0$ 
would contradict the condition $\gamma=1$ considered here. 

From the asymptotic expression for $\Gamma_{\sigma}({\Bf k})$ in 
Eq.~(\ref{e76}), making use of Eq.~(\ref{e90}), for ${\Bf k} \to 
{\cal S}_{{\sc f};\sigma}$ we obtain
\begin{equation}
\label{e92}
\Gamma_{\sigma}({\Bf k}) \sim
\frac{ -( a_{\sigma} +
U\, b_{\sigma}^{\mp} ) }
{ a_{\sigma} - U\, \Lambda_{\sigma}({\Bf k})\,
b_{\sigma}^{\mp} }, \;\;\; 
{\Bf k} \in \mbox{\rm FS}_{\sigma},\,
\overline{\mbox{\rm FS}}_{\sigma},
\end{equation}
where 
\begin{equation}
\label{e93} 
a_{\sigma} {:=} {\Bf a}({\Bf k}_{{\sc f};\sigma}) \cdot
{\hat{\Bf n}}({\Bf k}_{{\sc f};\sigma}), 
\end{equation}
\begin{equation}
\label{e94}
b_{\sigma}^{\mp} {:=} {\Bf b}_{\sigma}^{\mp} \cdot  
{\hat{\Bf n}}({\Bf k}_{{\sc f};\sigma}).
\end{equation}
Below we similarly employ the notation
\begin{equation}
\label{e95}
\Lambda_{\sigma}^{\mp} {:=} 
\Lambda_{\sigma}({\Bf k}_{{\sc f};\sigma}^{\mp}).
\end{equation}
The requirement with regard to the stability of the GS of the system 
under consideration implies the simultaneous satisfaction of the 
following two conditions: 
\begin{equation}
\label{e96}
b^{-}_{\sigma} > \frac{1}{U\, \Lambda_{\sigma}^{-} }\, 
a_{\sigma} \;\;\;\mbox{\rm (I)} \;\;\;\;\;\;\;\;\;\;\;
b^{+}_{\sigma} < -\frac{1}{U}\, a_{\sigma} 
\;\;\; \mbox{\rm (II)}
\end{equation}
which guarantee that the numerator and the denominator of the 
expression on the RHS of Eq.~(\ref{e76}) are both positive. In 
arriving at the above conditions, we have taken into account the 
facts that $U > 0$, $\Lambda_{\sigma}({\Bf k}) \ge 0$ (see 
Eq.~(\ref{e58})) and $a_{\sigma} \ge 0$. 

Solving Eq.~(\ref{e83}) for $\Lambda_{\sigma}^{\mp}$ (through
equating ${\Bf k}$ in $\Lambda_{\sigma}({\Bf k})$ with 
${\Bf k}_{{\sc f};\sigma}^{\mp}$), on the basis of the expression 
in Eq.~(\ref{e92}) we obtain the following quadratic equation 
\begin{equation}
\label{e97}
U\,b^{\mp}_{\sigma}\,
\big(\Lambda_{\sigma}^{\mp}\big)^2 - 
a_{\sigma}\, \Lambda_{\sigma}^{\mp} - 
(a_{\sigma}+U\, b^{\mp}_{\sigma}) =0.
\end{equation}
In view of the conditions in Eq.~(\ref{e96}) and the non-negativity of 
$\Lambda_{\sigma}({\Bf k})$, of the two solutions of Eq.~(\ref{e97}) 
for each of the vectors ${\Bf k} ={\Bf k}_{{\sc f};\sigma}^{-}$ and 
${\Bf k}={\Bf k}_{{\sc f};\sigma}^{+}$, only the following are acceptable
\begin{equation}
\label{e98}
\Lambda_{\sigma}^{\mp} =
1 + \frac{a_{\sigma}}
{U\, b^{\mp}_{\sigma} } \;\Ieq<> \; 1.
\end{equation}
From this and Eq.~(\ref{e85}) we obtain 
\begin{equation}
\label{e99}
{\sf n}_{\sigma}({\Bf k}_{{\sc f};\sigma}^{\mp})
= \frac{ a_{\sigma} + U\, b_{\sigma}^{\mp} }
{ a_{\sigma} + 2 U\, b_{\sigma}^{\mp} } \;\Ieq<>\; 
\frac{1}{2}.
\end{equation}
It is easily seen that for $a_{\sigma}=0$ (signifying ${\Bf k}_{{\sc f};
\sigma}$ as a saddle point, also referred to as a van Hove point [in 
regard to the density of the non-interacting single-particle states], 
of $\varepsilon_{\bF k}$), ${\sf n}_{\sigma}({\Bf k}_{{\sc f};
\sigma}^{\mp})=1/2$ and for $U\,\vert b_{\sigma}^{\mp}\vert \to \infty$, 
${\sf n}_{\sigma}({\Bf k}_{{\sc f};\sigma}^{\mp}) \to 1/2$, implying, 
through Eq.~(\ref{e64}), that $Z_{{\bF k}_{{\sc f};\sigma}} = 0$ in the 
former case and $Z_{{\bF k}_{{\sc f};\sigma}} \to 0$ in the latter. 
\footnote{\label{f5}
We point out that, for conventional energy dispersions, $a_{\sigma} 
\to 0$ for $n_{\sigma}\to 0$; that is, we similarly have
$Z_{{\bF k}_{{\sc f};\sigma}} \to 0$ in the low-density limit.}
Later, for FL metallic states (encompassing those which in spite of 
${\cal S}_{{\sc f};\sigma}^{(0)}\backslash {\cal S}_{{\sc f};\sigma}
\not=\emptyset$, share the basic characteristics of FLs) we deduce $U\, 
{\Bf b}_{\sigma}^{-} = \hbar\big(\lambda {\Bf v}_{{\sc f};\sigma}
-{\Bf v}_{{\sc f};\sigma}^{(0)}\big)$ (see Eq.~(\ref{e111}) below) and 
$U\, {\Bf b}_{\sigma}^{+} = -\hbar\lambda {\Bf v}_{{\sc f};\sigma}$, 
where $\lambda$ is a finite {\sl positive} constant. Thus for FLs 
$U\,\vert b_{\sigma}^{\mp}\vert$ are bounded even when $U\to\infty$, 
since for these systems $\|{\Bf v}_{{\sc f};\sigma}\|$ is finite. 
Evidently, this observation concerning $U\to\infty$ solely demonstrates 
the self-consistency of the FL theory within the framework of our 
considerations in this paper and does not rule out the 
possibility of the breakdown of a FL metallic state as $U$ is 
increased towards large values (for instance, through the change 
of $\gamma=1$ into $\gamma=\gamma_0 < 1$).

For completeness, we mention that quantum Monte Carlo results obtained
by Hlubina, {\sl et al.} \cite{HSG97}, corresponding to a $16\times 16$ 
square lattice (for $R {:=} 2t'/t = 0.94$, $U/t=2$ and $n=
n_{\rm v\sc h}$, where $t'$ stands for the hopping integral 
corresponding to next-nearest-neighbour sites and $n_{\rm v\sc h}$ 
for the van Hove density --- for definition, see above) ``suggest 
that the smearing of the momentum distribution function ${\sf n}
({\Bf k})$ does not increase close to the VH [van Hove] points.'' For 
the model energy dispersion $\varepsilon_{\bF k} = k_x k_y$ and the 
Fermi energy $\varepsilon_{\sc f}=0$ at the saddle-point ${\Bf k}
={\bf 0}$, Hlubina {\sl et al.} \cite{HSG97} reported a finite 
$Z_{\sc f}$ within the framework of the $t$-matrix-approximation 
(TMA) and further ${\rm Im}[\Sigma({\Bf k}={\bf 0};\varepsilon)] 
\sim -\varepsilon/\ln^2\vert\varepsilon\vert$ which, although 
unconventional for $d=2$, does {\sl not} imply breakdown of the 
quasiparticle picture (a fact already implicit in $Z_{\sc f} > 0$). 
The renormalization-group analysis by Gonz\'alez, {\sl et al.} 
\cite{GGV96,GGV97} for a model of two-dimensional electrons (closely 
related to the Hubbard Hamiltonian, involving quadratic energy 
dispersions close to the van Hove points), establishes a marginal-FL 
\cite{VLSRAR89} behaviour for the single-particle excitations in 
the vicinity of the van Hove points (see also \cite{IED96},
\cite{GGV96,GGV97} and \cite{IKK01}). The above opposing 
observations indicate that our rigorous finding relating 
$a_{\sigma}=0$ to the breakdown of the FL state at ${\Bf k}
={\Bf k}_{{\sc f};\sigma}$ cannot be viewed as being 
{\sl a priori} evident. 

Whether ${\sf n}_{\sigma}({\Bf k}_{{\sc f};\sigma})$, as presented 
in Eq.~(\ref{e65}), is equal to $1/2$, depends on the equality of
$1- {\sf n}_{\sigma}({\Bf k}_{{\sc f};\sigma}^{-})$ with 
${\sf n}_{\sigma}({\Bf k}_{{\sc f};\sigma}^{+})$; through some 
algebra we obtain
\begin{equation}
\label{e100}
1- {\sf n}_{\sigma}({\Bf k}_{{\sc f};\sigma}^{-})
= \frac{U\, b_{\sigma}^{-} }
{a_{\sigma} + 2 U\, b_{\sigma}^{-} },
\end{equation}
which cannot be unconditionally equal to ${\sf n}_{\sigma}
({\Bf k}_{{\sc f};\sigma}^{+})$, presented in Eq.~(\ref{e99}) above. 
For completeness, in contrast with case I considered above, in the 
present case and provided that $Z_{{\bF k}_{{\sc f};\sigma}}\not=0$, 
one can similar to the case of FLs (to be considered below) ascribe 
effective mass and (Fermi) velocity to low-lying single-particle
excitations of the system; however, in contrast with the case of FLs,
these characteristics are different for the excitations inside and 
outside the Fermi sea FS$_{\sigma}$. Consequently, for the present 
case, one can directly generalize, for instance, the semi-classical 
theory of metallic conduction (see Chapter 13 in \cite{AM81}). By 
doing so, one would obtain a generalized Drude formula for the
conductivity tensor which along its principal axes would involve two 
distinctive masses, corresponding to the single-particle excitations 
inside and outside FS$_{\sigma}$ close to ${\cal S}_{{\sc f};
\sigma}$. In this generalized Drude formula, the quantitative 
difference between the masses can be formally fully accounted for 
by introducing two distinctive relaxation times. Further, the 
discontinuity in the present case in the `Fermi velocity' across the 
Fermi surface ${\cal S}_{{\sc f};\sigma}$ gives rise to a 
temperature-independent contribution to the conductivity, or the 
associated relaxation times. These observations should be relevant 
to the physics of the cuprates in view of the appearance of two 
transport relaxation times in various transport coefficients of 
these materials (see \cite{PWA91,CST96,SS01}).

\vspace{0.4cm}
\noindent{11.2.2. \it Fermi-liquid metallic states}
\vspace{0.2cm}

We consider a metallic state as being a FL provided that the
following conditions hold \cite{BF99a}.
\begin{quote}
(A) $\Sigma_{\sigma}({\Bf k};\varepsilon)$, with ${\Bf k}
\in {\cal S}_{{\sc f};\sigma}$, is continuously differentiable 
with respect to $\varepsilon$ in a neighbourhood of $\varepsilon
=\varepsilon_{\sc f}$;
\end{quote}
\begin{quote}
(B) $\Sigma_{\sigma}({\Bf k};
\varepsilon_{\sc f})$ is continuously differentiable with respect 
to ${\Bf k}$ in a neighbourhood of ${\Bf k}_{{\sc f};\sigma}$, 
$\forall {\Bf k}_{{\sc f};\sigma} \in {\cal S}_{{\sc f};\sigma}$.
\end{quote}
Since ${\cal S}_{{\sc f};\sigma}^{(0)}$ does not feature in these 
conditions, the cases corresponding to ${\cal S}_{{\sc f};
\sigma}^{(0)}\backslash {\cal S}_{{\sc f};\sigma}\not=\emptyset$ 
are also accounted for here. We note in passing that satisfaction 
of condition (A) is {\sl sufficient} for $Z_{{\bF k}_{{\sc f};\sigma}}
\not=0$, so that metallic states with $Z_{{\bF k}_{{\sc f};\sigma}} 
\not=0$ are {\sl not} necessarily FLs.

For FL metallic states
\begin{eqnarray}
&&\left| \frac{{\rm Im}[\Sigma_{\sigma}({\Bf k};\varepsilon)]}
{{\rm Re}[\Sigma_{\sigma}({\Bf k};\varepsilon)] -
\Sigma_{\sigma}({\Bf k};\varepsilon_{\sc f})}\right| \sim
C_{\sigma}({\Bf k}) 
\vert \varepsilon-\varepsilon_{\sc f}\vert^{\alpha}
\left| \ln\vert\varepsilon-\varepsilon_{\sc f}\vert \right|^{\beta}
\nonumber
\end{eqnarray}
as $\varepsilon\to \varepsilon_{\sc f}$, where $C_{\sigma}({\Bf k}) 
> 0$, $\alpha > 0$ and $\vert\beta\vert \ge 0$. On account of this, 
neglecting ${\rm Im}[\Sigma_{\sigma}({\Bf k};\varepsilon)]$, the 
equation for the single-particle excitation energies, i.e. 
Eq.~(\ref{e12}), yields real-valued solutions; these, which we 
denote by $\varepsilon_{{\bF k};\sigma}$, describe the mean values 
of the peaks in $A_{\sigma}({\Bf k};\varepsilon)$ for $({\Bf k};
\varepsilon)$ close to $({\Bf k}_{{\sc f};\sigma};\varepsilon_{\sc f})$. 
Viewed from this perspective, for 
sufficiently small $\| {\Bf k}-{\Bf k}_{{\sc f};\sigma}\|$, on account
of Eqs.~(\ref{e12}), (\ref{e53}) and (\ref{e61}) for FLs we must have 
\begin{eqnarray}
\label{e101}
{\Bf\nabla}_{\bF k} \Big( \varepsilon_{\bF k}
+U\, \frac{\beta_{{\bF k};\sigma}^{\Ieq><} }
{ \nu_{\sigma}^{\Ieq><}({\Bf k}) } \Big)
&\sim& \lambda\, {\Bf\nabla}_{\bF k} \Big(
\varepsilon_{\bF k} + \hbar 
\Sigma_{\sigma}({\Bf k};\varepsilon_{{\bF k};\sigma}) \Big), 
\nonumber\\
& &\;\;\;\;\;\;\;\;\;\;\;\;\;
\varepsilon_{{\bF k};\sigma} \, \Ieq>< \, \mu,
\end{eqnarray} 
where $\lambda$ is a {\sl positive} constant which on general
grounds we expect to be of the order of unity (see the paragraph
following Eq.~(\ref{e112}) below). On the other hand, for FL 
metallic states we have (in consequence of the conditions (A) 
and (B) introduced above)
\begin{equation}
\label{e102}
{\Bf\nabla}_{\bF k}\Sigma_{\sigma}({\Bf k};
\varepsilon_{{\bF k};\sigma}) \sim 
\big({\Bf v}_{{\sc f};\sigma} - {\Bf v}_{{\sc f};\sigma}^{(0)} \big),
\;\;\; \mbox{\rm for}\;\;\;\| {\Bf k}-{\Bf k}_{{\sc f};\sigma} \| \to 0,
\end{equation}
where ${\Bf v}_{{\sc f};\sigma}^{(0)}$ is defined in Eq.~(\ref{e73}), and
\begin{eqnarray}
\label{e103}
{\Bf v}_{{\sc f};\sigma} &{:=}&
\frac{1}{\hbar}\left. {\Bf\nabla}_{\bF k} 
\varepsilon_{{\bF k};\sigma}\right|_{{\bF k}={\bF k}_{{\sc f};\sigma}}
\nonumber\\
&\equiv&
Z_{{\bF k}_{{\sc f};\sigma}} 
\Big( {\Bf v}_{{\sc f};\sigma}^{(0)}
+ {\Bf\nabla}_{\bF k} \left. 
\Sigma_{\sigma}({\Bf k};\varepsilon_{\sc f})\right|_{{\bF k}
= {\bF k}_{{\sc f};\sigma}}\Big),
\end{eqnarray}
in which $Z_{{\bF k}_{{\sc f};\sigma}}$ is defined in Eq.~(\ref{e64}) 
and for which we have the following alternative expression:
\begin{equation}
\label{e104}
Z_{{\bF k}_{{\sc f};\sigma}} = \Big( 1 - \hbar \left.
{\rm d}\Sigma_{\sigma}({\Bf k}_{{\sc f};\sigma};\varepsilon)/
{\rm d}\varepsilon\right|_{\varepsilon
=\varepsilon_{\sc f}}\Big)^{-1}.
\end{equation}

\begin{figure}[t!]
\protect
\centerline{
\psfig{figure=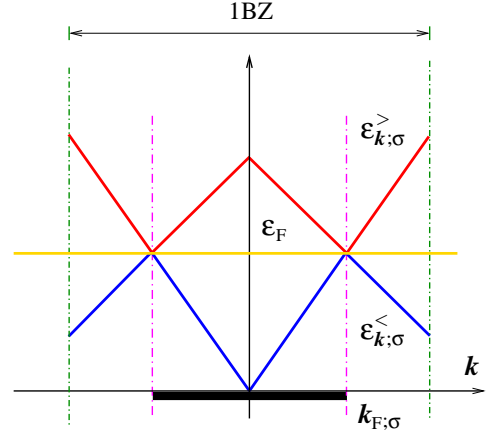,width=2.5in} }
\vskip 5pt
\caption{\label{fi3} \sf
Schematic forms of 
$\varepsilon_{{\bF k};\sigma}^{<}$
and $\varepsilon_{{\bF k};\sigma}^{>}$ 
for FL metallic states. Note that 
(see Eqs.~(\protect\ref{e105}) and (\protect\ref{e106}))
$\varepsilon_{{\bF k};\sigma}^{<}$ 
($\varepsilon_{{\bF k};\sigma}^{>}$)
smoothly goes over into $\varepsilon_{{\bF k};\sigma}^{>}$ 
($\varepsilon_{{\bF k};\sigma}^{<}$) for ${\Bf k}$ transposed from 
inside to outside FS$_{\sigma}$. } 
\end{figure}

The assumption with regard to the continuous differentiability of
$\varepsilon_{{\bF k};\sigma}$ in a neighbourhood of 
${\cal S}_{{\sc f};\sigma}$ implies that we have (Fig.~\ref{fi3})
\begin{eqnarray}
\label{e105}
&&\left. {\Bf\nabla}_{\bF k}
\varepsilon_{{\bF k};\sigma}^{<}\right|_{{\bF k}
={\bF k}_{{\sc f};\sigma}^-} =
\left. {\Bf\nabla}_{\bF k}
\varepsilon_{{\bF k};\sigma}^{>}\right|_{{\bF k}
={\bF k}_{{\sc f};\sigma}^+} \nonumber \\
&&\;\;\;\;\;\;\;\;\;\;\;\;\;\;\;\;\;\;\;
\Longleftrightarrow\, 
{\Bf a}_{\sigma} + U {\Bf b}_{\sigma}^- 
= {\Bf a}_{\sigma} - U \Lambda_{\sigma}^+\, {\Bf b}_{\sigma}^+. 
\end{eqnarray}
As we demonstrate below, these results are equivalent to (see 
Fig.~\ref{fi3})
\begin{eqnarray}
\label{e106}
&&\left. {\Bf\nabla}_{\bF k}
\varepsilon_{{\bF k};\sigma}^{>}\right|_{{\bF k}
={\bF k}_{{\sc f};\sigma}^-} =
\left. {\Bf\nabla}_{\bF k}
\varepsilon_{{\bF k};\sigma}^{<}\right|_{{\bF k}
={\bF k}_{{\sc f};\sigma}^+} \nonumber \\
&&\;\;\;\;\;\;\;\;\;\;\;\;\;\;\;\;\;\;\;
\Longleftrightarrow\,
{\Bf a}_{\sigma} + U {\Bf b}_{\sigma}^+ 
= {\Bf a}_{\sigma} - U \Lambda_{\sigma}^-\, {\Bf b}_{\sigma}^-.
\end{eqnarray}
\noindent
To demonstrate the equivalence of the results in Eqs.~(\ref{e105}) and 
(\ref{e106}), we first consider Eq.~(\ref{e105}) from which we have 
${\Bf b}_{\sigma}^- = -\Lambda_{\sigma}^+\, {\Bf b}_{\sigma}^+$ which 
makes explicit that ${\Bf b}_{\sigma}^-$ and ${\Bf b}_{\sigma}^+$ are 
collinear and point in opposite directions. Making use of the 
expression for $\Lambda_{\sigma}^+$ as presented in Eq.~(\ref{e98}), 
one readily obtains
\begin{equation}
\label{e107}
{\Bf b}_{\sigma}^+ = - {\Bf b}_{\sigma}^- - \frac{1}{U}\,
{\Bf a}_{\sigma}.
\end{equation}
This result, which implies that $b_{\sigma}^+ = -b_{\sigma}^- -
a_{\sigma}/U$, in combination with Eq.~(\ref{e98}) yields
\begin{equation}
\label{e108}
\Lambda_{\sigma}^+ =\frac{1}{\Lambda_{\sigma}^-}
\equiv \frac{U\, b_{\sigma}^-}{a_{\sigma} + U\, b_{\sigma}^-}.
\end{equation}
From this and ${\Bf b}_{\sigma}^- = -\Lambda_{\sigma}^+\, 
{\Bf b}_{\sigma}^+$ we deduce ${\Bf b}_{\sigma}^+ = 
-\Lambda_{\sigma}^-\, {\Bf b}_{\sigma}^-$, or
${\Bf a}_{\sigma} + U {\Bf b}_{\sigma}^+ = {\Bf a}_{\sigma} - 
\Lambda_{\sigma}^-\, U {\Bf b}_{\sigma}^-$, which is Eq.~(\ref{e106}). 

From Eqs.~(\ref{e85}) and (\ref{e108}) we obtain
\begin{equation}
\label{e109}
{\sf n}_{\sigma}({\Bf k}_{{\sc f};\sigma}^+)
= \frac{U\, b_{\sigma}^-}{a_{\sigma} + 2 U\, b_{\sigma}^-}
\equiv 1 - {\sf n}_{\sigma}({\Bf k}_{{\sc f};\sigma}^-).
\end{equation}
Thus by Eq.~(\ref{e65}) we have
\begin{equation}
\label{e110}
{\sf n}_{\sigma}({\Bf k}_{{\sc f};\sigma}) = \frac{1}{2}.
\end{equation}
This result is significant in that it unequivocally demonstrates that 
a uniform metallic GS of the Hubbard Hamiltonian whose zero-temperature 
limit of ${\sf n}_{\sigma}({\Bf k})$ is not equal to $1/2$ for ${\Bf k} 
\in {\cal S}_{{\sc f};\sigma}$, {\sl cannot} be a FL. Interestingly, 
the finite-temperature ($T=0.015 t/k_{\sc b}$) numerical calculations 
based on the fluctuation-exchange, FLEX, approximation to the self-energy 
operator of the single-band Hubbard Hamiltonian on a square lattice 
for $d=2$ with $U/t=8$ \cite{SH91} indicate ${\sf n}_{\sigma}({\Bf k})$ 
to deviate from $1/2$ for ${\Bf k} \in {\cal S}_{{\sc f};\sigma}$; for 
$n_{\sigma}=n_{\bar\sigma}=0.265$ and ${\Bf k}_{{\sc f};\sigma} = \pi 
(0.61024\dots,0)$ (in units of the inverse lattice constant), 
${\sf n}_{\sigma}({\Bf k}_{{\sc f};\sigma}) \approx 0.38$ (see Fig.~4 
in \cite{SH91}). By considering the \tJ Hamiltonian as the 
strong-coupling limit of the Hubbard Hamiltonian (see footnote
\ref{f0}), in which limit the small exchange integral $J$ is formally 
equal to $4t^2/U$ \cite{CSO78,GJR87}, accurate numerical results for 
${\sf n}_{\sigma}({\Bf k})$ pertaining to uniform GSs of the \tJ 
Hamiltonian (deduced from a twelfth-order expansion in the inverse 
temperature \cite{SG92} and corresponding to $T = 0.2 J/k_{\sc b}$) 
for $J/t = 0.5$ and $1.0$ \cite{PLS98a} and specifically $J/t = 0.4$ 
\cite{PLS98b} ($J/t \approx 0.4$ formally corresponds to $U/t \approx 
10.0$) confirm our above conclusion that the underlying uniform 
metallic GSs {\sl cannot} be FL (see Appendix for some details 
concerning aspects of the ${\sf n}_{\sigma}({\Bf k})$ associated with 
the \tJ Hamiltonian). As our finding in Eq.~(\ref{e24}) makes evident, 
the interpretation provided by the authors \cite{PLS98a,PLS98b}, namely 
that the Luttinger theorem \cite{L60,LW60} concerning the number of 
${\Bf k}$ points enclosed by the ${\cal S}_{{\sc f};\sigma}$ associated 
with the underlying metallic GS, breaks down, {\sl cannot} be the 
appropriate interpretation. We note in passing that the Fermi surface 
as depicted in Fig.~1 of \cite{PLS98b} is the locus of the ${\Bf k}$ 
points for which ${\sf n}_{\sigma}({\Bf k}) =1/2$ \cite{REA95}. It is 
important to point out that, since according to the numerical results 
in \cite{PLS98a,PLS98b} ${\sf n}_{\sigma}({\Bf k}) \not=1/2$ for 
${\Bf k} \in {\cal S}_{{\sc f};\sigma}$, it is excluded that 
$Z_{\bF k}=0$ for ${\Bf k}\in {\cal S}_{{\sc f};\sigma}$, so that 
the underlying states must fail to be FL solely on account of the 
failure of condition (B) presented above. 

\vspace{2.5mm}

In connection with the above observations, we mention that the locus 
of the ${\Bf k}$ points for which ${\sf n}_{\sigma}({\Bf k})=1/2$, 
with ${\sf n}_{\sigma}({\Bf k})$ the extrapolation (based on a finite
mesh of ${\Bf k}$ points) of the finite-temperature 
quantum-Monte-Carlo results by Moreo, {\sl et al.} \cite{MSSWB90}, 
pertaining to the Hubbard Hamiltonian on a $16\times 16$ square 
lattice with $U/t=4$ (at $T = t/(6 k_{\sc b})$ and for $n_{\sigma}
=n_{\bar\sigma}=0.435$) almost identically coincides with 
${\cal S}_{{\sc f};\sigma}^{(0)}$. As is evident from our above 
considerations, this result is not sufficient (although it is 
necessary) for the underlying metallic state to be identified as a 
FL. From the available numerical data we are not capable of 
deducing whether the FL metallic state breaks down for $U/t > 
U_{\rm c}/t$, with $U_{\rm c}$ a finite (positive) constant, although 
in view of the above-mentioned Monte Carlo results \cite{MSSWB90} it 
is tempting to believe that $U_{\rm c}/t\,\Ieq{\sim}{>}\, 4$. This 
in particular in view of the fact that, for $U/t=4$ (and $T=0.05 
t/k_{\sc b}$) the ${\sf n}_{\sigma}({\Bf k})$ as obtained within 
the framework of the FLEX approximation \cite{BW91} does {\sl not} 
show any noticeable asymmetry with respect to $1/2$ for ${\Bf k} = 
{\Bf k}_{{\sc f};\sigma}^{\mp}$ (see Fig.~13 in \cite{BW91}), in stark 
contrast with the ${\sf n}_{\sigma}({\Bf k})$ corresponding to $U/t=8$ 
\cite{SH91}, discussed above.

From Eqs.~(\ref{e101}), (\ref{e102}) and (\ref{e105}) and some simple 
algebra, we obtain
\begin{equation}
\label{e111}
{\Bf v}_{{\sc f};\sigma} = \frac{1}{\lambda}\,
\Big({\Bf v}_{{\sc f};\sigma}^{(0)} + \frac{1}{\hbar} U\, 
{\Bf b}_{\sigma}^- \Big)
\equiv \frac{1}{\lambda}\,
\frac{ {\sf n}_{\sigma}({\Bf k}_{{\sc f};\sigma}^-)}
{2 {\sf n}_{\sigma}({\Bf k}_{{\sc f};\sigma}^-) -1}\,
{\Bf v}_{{\sc f};\sigma}^{(0)}.
\end{equation}
As we have indicated earlier, the three vectors in Eq.~(\ref{e111}) 
are collinear and point in the direction of the outward normal vector 
to ${\cal S}_{{\sc f};\sigma}$ at ${\Bf k}_{{\sc f};\sigma}$. Thus 
Eq.~(\ref{e111}) can be identically written as $v_{{\sc f};\sigma} = 
\big(v_{{\sc f};\sigma}^{(0)} + U\, b_{\sigma}^-/\hbar\big)/\lambda$,
from which it follows that, by the requirement of the stability of the 
GS of the system, $\lambda > 0$ must hold (recall that through
Eq.~(\ref{e96}) we have $b_{\sigma}^- > 0$ for $U > 0$). From 
Eq.~(\ref{e111}) we deduce that
\begin{equation}
\label{e112}
v_{{\sc f};\sigma}\,\Ieq<>\, v_{{\sc f};\sigma}^{(0)}\;\;
\Longleftrightarrow \;\;
\lambda\, \Ieq>< \, 
\frac{{\sf n}_{\sigma}({\Bf k}_{{\sc f};\sigma}^-)}
{2 {\sf n}_{\sigma}({\Bf k}_{{\sc f};\sigma}^-) -1}.
\end{equation}
For isotropic FLs, one has $v_{{\sc f};\sigma} = 
(m_{\rm e}/m_{\sigma}^*)\, v_{{\sc f};\sigma}^{(0)}$, whereby 
$v_{{\sc f};\sigma} \,\Ieq<>\, v_{{\sc f};\sigma}^{(0)}$ implies 
that $m_{\sigma}^* \,\Ieq><\, m_{\rm e}$; here $m_{\rm e}$ stands 
for the bare electron mass and $m_{\sigma}^*$ for the effective 
mass of the quasiparticles with spin index $\sigma$. For isotropic 
FLs we thus have $\lambda = (m_{\sigma}^*/m_{\rm e}) 
{\sf n}_{\sigma}(k_{{\sc f};\sigma}^-)/\big(2 
{\sf n}_{\sigma}(k_{{\sc f};\sigma}^-) -1 \big)$.

It is interesting to note that, for $U b_{\sigma}^{-} \to 0$,
Eq.~(\ref{e109}) yields ${\sf n}_{\sigma}({\Bf k}_{{\sc f};
\sigma}^+) \downarrow 0$ and ${\sf n}_{\sigma}({\Bf k}_{{\sc f};
\sigma}^-) \uparrow 1$. We observe that ${\sf n}_{\sigma}
({\Bf k}_{{\sc f};\sigma}^{\pm})$ as calculated in this Section 
have the correct limits for the diminishing strength of 
interaction, in conformity with the fact that non-interacting 
metallic states are indeed (ideal) FLs. It follows that 
${\sf n}_{\sigma}({\Bf k}_{{\sc f};\sigma}^-)/\big(2 {\sf n}_{\sigma}
({\Bf k}_{{\sc f};\sigma}^-)-1\big) \to 1$ for $U b_{\sigma}^{-} 
\to 0$ so that, in view of the requirement that in this limit
${\Bf v}_{{\sc f};\sigma} \to {\Bf v}_{{\sc f};\sigma}^{(0)}$, 
Eq.~(\ref{e111}) is seen to imply the expected result that 
$\lambda\to 1$ for $U b_{\sigma}^{-} \to 0$; it is {\sl a priori}
not excluded that at least $\lambda\approx 1$ for {\sl all} 
values of $U b_{\sigma}^{-}$ for which the metallic state under 
consideration is a FL. We note in passing that Eqs.~(\ref{e105}) 
and (\ref{e106}) imply that $\varepsilon_{{\bF k};\sigma}^{<}$ 
and $\varepsilon_{{\bF k};\sigma}^{>}$ form a Dirac-type spectrum 
for ${\Bf k}$ approaching ${\cal S}_{{\sc f};\sigma}$ (see
Fig.~\ref{fi3}).

\subsubsection*{{\rm 11.3.}~\it Case III: $\; \gamma > 1$}

In this case, from Eqs.~(\ref{e74}) and (\ref{e75}) it follows that, 
for ${\Bf k}\to {\cal S}_{{\sc f};\sigma}$, $\mu - 
\varepsilon_{{\bF k};\sigma}^{<} \sim
\varepsilon_{{\bF k}_{{\sc f};\sigma}} - \varepsilon_{\bF k}$ and 
$\varepsilon_{{\bF k};\sigma}^{>} -\mu \sim \varepsilon_{\bF k} 
-\varepsilon_{{\bF k}_{{\sc f};\sigma}}$. Both of these results 
imply instability of the GS of the system, as they imply 
$\varepsilon_{{\bF k};\sigma}^{<} > \mu$ for ${\Bf k} \in 
\ol{\rm FS}_{\sigma}$ and $\varepsilon_{{\bF k};\sigma}^{>} < \mu$ 
for ${\Bf k} \in {\rm FS}_{\sigma}$. Consequently, $\gamma > 1$ 
is {\sl not} permissible.

\subsection*{\S~12. \sc Concluding remarks}

In conclusion, our considerations in this work make explicit the 
considerable significance of the non-interacting energy dispersion 
$\varepsilon_{\bF k}$, in terms of which the single-band Hubbard 
Hamiltonian $\wh{\cal H}$ is defined, in accurately describing the 
low-energy single-particle excitations of the associated correlated 
metallic systems; not only is the shape of ${\cal S}_{{\sc f};\sigma}$ 
determined by $\varepsilon_{\bF k}$ (leaving aside the dependence 
on $N_{\sigma}$, and thus on $U$, of $\epsilon_{\sigma}$ through 
Eq.~(\ref{e21})), but also can the suppression to zero of the 
quasiparticle weight $Z_{{\bF k}_{{\sc f};\sigma}}$ at 
${\Bf k}_{{\sc f};\sigma} \in {\cal S}_{{\sc f};\sigma}$ be brought 
about ({\sl not} exclusively) by the choice of $\varepsilon_{\bF k}$ 
in such a way that, for a given $n_{\sigma}$, ${\Bf k}_{{\sc f};\sigma}$ 
coincides with a saddle point of $\varepsilon_{\bF k}$. Although our 
theoretical findings combined with some available numerical results 
concerning the Hubbard and the \tJ Hamiltonian reveal that on a 
two-dimensional square lattice, and at least for $U/t\,\Ieq{\sim}{>}\,
8$, the metallic GSs of the single-band Hubbard Hamiltonian are 
{\sl not} FL, the NFL states at issue are in general (that is, for 
a rigid band energy dispersion $\varepsilon_{\bF k}$ described in 
terms of nearest-neighbour and next-nearest-neighbour hopping 
integrals, $t$ and $t'$ respectively) neither pure marginal-FL nor 
pure Luttinger-liquid (in spite of the assumed homogeneity of the 
GS, depending on $\varepsilon_{\bF k}$ and the band fillings 
$n_{\sigma}$ and $n_{\bar\sigma}$, the low-lying single-particle 
excitations can be strongly anisotropic in the momentum space); the 
latter states are specifically characterized by single-particle 
spectral functions which lack a coherent low-energy quasiparticle 
peak on ${\cal S}_{{\sc f};\sigma}$ and are variously identified 
with the normal metallic states of the cuprate superconductors. 
Assuming the sufficiency of the single-band Hubbard Hamiltonian for 
describing the low-lying single-particle excitations of the cuprate 
superconductors in their normal metallic states, our findings imply 
that further considerations based on more general energy dispersions 
than those expressed in terms of $t$ and $t'$ are paramount; our 
considerations make explicit that the functional form of such energy 
dispersions may even be required to depend on the band filling $n$, 
that is, a rigid $\varepsilon_{\bF k}$ may in general not suffice. 
The significance of the $t'$, in particular in the underdoped 
(in connection with the pseudogap phenomenon in cuprates) and overdoped 
(in connection with the occurrence of ferromagnetism and strong 
ferromagnetic fluctuations) regions of the phase diagram of the 
single-band Hubbard Hamiltonian for $d=2$ has been already recognized.

\vspace{1mm}
\subsection*{\sc Acknowlegements}
With pleasure I thank
Professor Ganapathy Baskaran,
Dr Peter Horsch,
Professor William Putikka,
Professor Rajiv Singh and
Professor Philip Stamp
for discussion and provision of information concerning
some computational aspects specific to the \tJ model,
and Spinoza Institute for hospitality and support.

\vspace{3mm}

\begin{appendix}

\section*{A.~\sc Constrained operators and the associated
momentum distribution function}

The \tJ Hamiltonian is defined in terms of constrained (or projected) 
site operators $\{ {\hat a}_{i;\sigma} \}$ which are expressed in 
terms of the canonical site operators $\{ {\hat c}_{i;\sigma} \}$ 
as follows \cite{HL67,CSO78,GJR87} 
\begin{equation}
\label{ea1}
{\hat a}_{i;\sigma} {:=} {\hat c}_{i;\sigma} 
(1 - {\hat n}_{i;\bar\sigma}). 
\end{equation}
With 
\begin{equation}
\label{ea2}
{\hat a}_{{\bF k};\sigma} {:=}
\frac{1}{N_{\sc l}^{1/2}} \sum_i
{\hat a}_{i;\sigma}\, {\rm e}^{-i {\bF k}\cdot {\bF R}_i },
\end{equation}
and $\vert\Upsilon_{N}\rangle$ an $N$-particle normalized state, 
with $N=N_{\sigma}+N_{\bar\sigma}$, we define the following momentum 
distribution function
\begin{equation}
\label{ea3}
\tilde{\sf n}_{\sigma}({\Bf k}) 
{:=} \langle\Upsilon_{N}\vert
{\hat a}_{{\bF k};\sigma}^{\dag}
{\hat a}_{{\bF k};\sigma} \vert\Upsilon_{N}\rangle.
\end{equation}
In this appendix we investigate the relationship between 
$\tilde{\sf n}_{\sigma}({\Bf k})$ and the `conventional' 
momentum distribution function ${\sf n}_{\sigma}({\Bf k})$, 
defined as
\begin{equation}
\label{ea4}
{\sf n}_{\sigma}({\Bf k}) {:=}
\langle\Upsilon_{N}\vert
{\hat c}_{{\bF k};\sigma}^{\dag}
{\hat c}_{{\bF k};\sigma} \vert\Upsilon_{N}\rangle,
\end{equation}
by assuming $\vert\Upsilon_{N}\rangle$ to be a {\sl variational}
Ansatz for the GS $\vert\Psi_{N;0}\rangle$ of the \tJ Hamiltonian
in the subspace of the $N$-particle Hilbert space where site 
double occupancy is excluded, that is
${\hat c}_{i;\bar\sigma} {\hat c}_{i;\sigma} \vert\Psi_{N;0}\rangle
=0$, $\forall i$; in contrast, 
${\hat c}_{i;\bar\sigma} {\hat c}_{i;\sigma} \vert\Upsilon_{N}\rangle
\not=0$, for some or all $i$.

Making use of the commutation relation
\begin{equation}
\label{ea5}
\big[ {\hat c}_{i;\sigma}, {\hat n}_{j;\sigma'}\big]_{-}
=\delta_{i,j} \delta_{\sigma,\sigma'}\,
{\hat c}_{i;\sigma},
\end{equation}
we obtain
\begin{equation}
\label{ea6}
{\hat a}_{i;\sigma}^{\dag} 
{\hat a}_{j;\sigma} = 
{\hat c}_{i;\sigma}^{\dag} 
{\hat c}_{j;\sigma} + {\hat h}_{i,j;\sigma},
\end{equation}
where
\begin{equation}
\label{ea7}
{\hat h}_{i,j;\sigma} {:=}
\big( {\hat n}_{i;\bar\sigma} {\hat n}_{j;\bar\sigma}
- {\hat n}_{i;\bar\sigma} - {\hat n}_{j;\bar\sigma} \big)\,
{\hat c}_{i;\sigma}^{\dag} {\hat c}_{j;\sigma}.
\end{equation}
With
\begin{equation}
\label{ea8}
{\sf n}_{\sigma}({\Bf k}) 
= \frac{1}{N_{\sc l}} \sum_{i,j}
\langle\Upsilon_{N}\vert {\hat c}_{i;\sigma}^{\dag}
{\hat c}_{j;\sigma} \vert\Upsilon_{N}\rangle\,
{\rm e}^{i {\bF k}\cdot ({\bF R}_i - {\bF R}_j) },
\end{equation}
and similarly for $\tilde{\sf n}_{\sigma}({\Bf k})$,
from Eq.~(\ref{ea6}) it follows that
\begin{equation}
\label{ea9}
\tilde{\sf n}_{\sigma}({\Bf k}) =
{\sf n}_{\sigma}({\Bf k}) + \delta {\sf n}_{\sigma}({\Bf k}),
\end{equation}
where $\delta {\sf n}_{\sigma}({\Bf k})$ is related to
$\langle\Upsilon_{N}\vert {\hat h}_{i,j;\sigma}
\vert\Upsilon_{N}\rangle$ according to an expression similar 
to that in Eq.~(\ref{ea8}).

Making use of the canonical anticommutation relations for
${\hat c}_{i;\sigma}$ and ${\hat c}_{i;\sigma}^{\dag}$,
one readily obtains the following normal-ordered expression:
\begin{eqnarray}
\label{ea10}
&&\langle\Upsilon_{N}\vert {\hat h}_{i,j;\sigma} 
\vert\Upsilon_{N}\rangle \nonumber\\
&&\;\;\;\;\;\;\;\;\;
= -\langle\Upsilon_{N}\vert  
{\hat c}_{i;\bar\sigma}^{\dag}
{\hat c}_{i;\sigma}^{\dag}
{\hat c}_{i;\bar\sigma}
{\hat c}_{i;\sigma}\vert\Upsilon_{N}\rangle \,\delta_{i,j} 
\nonumber\\
&&\;\;\;\;\;\;\;\;\;\;\;\;\;
+\langle\Upsilon_{N}\vert  
{\hat c}_{i;\bar\sigma}^{\dag}
{\hat c}_{i;\sigma}^{\dag}
{\hat c}_{i;\bar\sigma}
{\hat c}_{j;\sigma}\vert\Upsilon_{N}\rangle 
\nonumber\\
&&\;\;\;\;\;\;\;\;\;\;\;\;\;
+\langle\Upsilon_{N}\vert  
{\hat c}_{j;\bar\sigma}^{\dag}
{\hat c}_{i;\sigma}^{\dag}
{\hat c}_{j;\bar\sigma}
{\hat c}_{j;\sigma}\vert\Upsilon_{N}\rangle 
\nonumber\\
&&\;\;\;\;\;\;\;\;\;\;\;\;\;
-\langle\Upsilon_{N}\vert  
{\hat c}_{i;\bar\sigma}^{\dag}
{\hat c}_{j;\bar\sigma}^{\dag}
{\hat c}_{i;\sigma}^{\dag}
{\hat c}_{i;\bar\sigma}
{\hat c}_{j;\bar\sigma}
{\hat c}_{j;\sigma}\vert\Upsilon_{N}\rangle. 
\end{eqnarray}
It is directly verified that, in the event that ${\hat c}_{i;\bar\sigma} 
{\hat c}_{i;\sigma} \vert\Upsilon_{N}\rangle = 0$, $\forall i$, 
the RHS of Eq.~(\ref{ea10}) is, as expected, vanishing. One further 
immediately observes that
\begin{eqnarray}
\label{ea11}
&&\langle\Upsilon_{N}\vert {\hat h}_{i,i;\sigma} 
\vert\Upsilon_{N}\rangle =
-\langle\Upsilon_{N}\vert  
{\hat c}_{i;\sigma}^{\dag}
{\hat c}_{i;\bar\sigma}^{\dag}
{\hat c}_{i;\bar\sigma}
{\hat c}_{i;\sigma}\vert\Upsilon_{N}\rangle 
\nonumber\\
&&\;\;\;\;\;\;\;\;\;\;\;\;\;\;
\equiv -\sum_s \left| \langle\Upsilon_{N-2;s}\vert 
{\hat c}_{i;\bar\sigma}
{\hat c}_{i;\sigma}\vert\Upsilon_{N}\rangle \right|^2 \le 0,
\end{eqnarray}
the equality sign applying for {\sl any} $i$ for which 
${\hat c}_{i;\bar\sigma} 
{\hat c}_{i;\sigma}\vert\Upsilon_{N}\rangle = 0$.
It follows that
\begin{equation}
\label{ea12}
\sum_{\bF k} \delta {\sf n}_{\sigma}({\Bf k}) \le 0,
\end{equation}
or equivalently (see Eq.~(\ref{ea9}))
\begin{equation}
\label{ea13}
\frac{1}{N_{\sc l}} \sum_{\bF k} \tilde{\sf n}_{\sigma}({\Bf k})
\le \frac{1}{N_{\sc l}} \sum_{\bF k} {\sf n}_{\sigma}({\Bf k})
= n_{\sigma} \equiv \frac{N_{\sigma}}{N_{\sc l}}.
\end{equation}
The equality signs in Eqs.~(\ref{ea12}) and (\ref{ea13}) apply in 
cases where ${\hat c}_{i;\bar\sigma} {\hat c}_{i;\sigma} 
\vert\Upsilon_{N}\rangle = 0$, $\forall i$.

In order to gain insight into the behaviour of 
$\delta {\sf n}_{\sigma}({\Bf k})$ and the amount that the sum on the 
left-hand side of Eq.~(\ref{ea13}) can deviate from $n_{\sigma}$, we 
now consider the case where $\vert\Upsilon_{N}\rangle$ is replaced 
by the normalized uncorrelated $N$-particle GS of $\wh{\cal H}_0$ 
(see Eq.~(\ref{e19})), which we denote by $\vert\Phi_{N;0}\rangle$, 
approximating the GS $\vert\Psi_{N;0}\rangle$. Following some 
algebraic manipulations, the details of which we shall present 
elsewhere, we obtain
\begin{eqnarray}
\label{ea14}
&&\langle\Phi_{N;0}\vert
{\hat c}_{i;\bar\sigma}^{\dag}
{\hat c}_{i;\sigma}^{\dag}
{\hat c}_{i;\bar\sigma}
{\hat c}_{j;\sigma} \vert\Phi_{N;0}\rangle 
=\langle\Phi_{N;0}\vert
{\hat c}_{j;\bar\sigma}^{\dag}
{\hat c}_{i;\sigma}^{\dag}
{\hat c}_{j;\bar\sigma}
{\hat c}_{j;\sigma} \vert\Phi_{N;0}\rangle 
\nonumber\\
&&\;\;\;\;\;\;\;\;\;\;\;\;\;\;\;\;
= \frac{1}{N_{\sc l}} \sum_{\bF k}
{\rm e}^{-i {\bF k}\cdot ({\bF R}_i-{\bF R}_j)}\,
\big[ -n_{\bar\sigma} {\sf n}_{\sigma}^{(0)}({\Bf k})\big],
\end{eqnarray}
\begin{eqnarray}
\label{ea15}
&&\langle\Phi_{N;0}\vert
{\hat c}_{i;\bar\sigma}^{\dag}
{\hat c}_{j;\bar\sigma}^{\dag}
{\hat c}_{i;\sigma}^{\dag}
{\hat c}_{i;\bar\sigma}
{\hat c}_{j;\bar\sigma} 
{\hat c}_{j;\sigma} 
\vert\Phi_{N;0}\rangle 
=\frac{1}{N_{\sc l}} \sum_{\bF k}
{\rm e}^{-i {\bF k}\cdot ({\bF R}_i-{\bF R}_j)}
\nonumber\\
&&\;\;\;\;\;
\times\big[\frac{1}{N_{\sc l}^2} \sum_{{\bF k}_1,{\bF k}_2}
{\sf n}_{\bar\sigma}^{(0)}({\Bf k}_1)
{\sf n}_{\bar\sigma}^{(0)}({\Bf k}_2)
{\sf n}_{\sigma}^{(0)}({\Bf k}-{\Bf k}_1+{\Bf k}_2) \nonumber\\
&&\;\;\;\;\;\;\;\;\;\;\;\;\;\;\;\;\;\;\;\;\;\;\;\;\;\;\;\;\;\;\;
\;\;\;\;\;\;\;\;\;\;\;\;\;\;\;\;\;\;\;\;\;\;\;\;\;\;\;
-n_{\bar\sigma}^2 {\sf n}_{\sigma}^{(0)}({\Bf k}) \big],
\end{eqnarray}
where ${\sf n}_{\sigma}^{(0)}({\Bf k}) {:=} \langle\Phi_{N;0}\vert 
{\hat c}_{{\bF k};\sigma}^{\dag} {\hat c}_{{\bF k};\sigma}
\vert\Phi_{N;0}\rangle$ is equal to the unit step function, unity for 
${\Bf k}\in {\rm FS}_{\sigma}^{(0)}$ and zero for ${\Bf k}\in 
\overline{\rm FS}_{\sigma}^{(0)}$. It is readily verified that for 
$i=j$ both sides of Eq.~(\ref{ea15}) are vanishing, exactly as in the 
case corresponding to the correlated GS $\vert\Psi_{N;0}\rangle$ for 
which we naturally also have ${\hat c}_{i;\bar\sigma} 
{\hat c}_{i;\bar\sigma} \vert\Psi_{N;0}\rangle = 0$, $\forall i$
(note the spin indices). Making use of
\begin{equation}
\label{ea16}
\frac{1}{N_{\sc l}} \sum_{\bF k}
{\rm e}^{-i {\bF k}\cdot ({\bF R}_i-{\bF R}_j)} = 
\delta_{i,j},
\end{equation}
for $\tilde{\sf n}_{\sigma}^{(0)}({\Bf k}) {:=} \langle\Phi_{N;0}\vert
{\hat a}_{{\bF k};\sigma}^{\dag} {\hat a}_{{\bF k};\sigma}
\vert\Phi_{N;0}\rangle$ from Eqs.~(\ref{ea9}), (\ref{ea10}), 
(\ref{ea14}) and (\ref{ea15}) we obtain
\begin{eqnarray}
\label{ea17}
&&\tilde{\sf n}_{\sigma}^{(0)}({\Bf k})
= {\sf n}_{\sigma}^{(0)}({\Bf k})
+ n_{\sigma} n_{\bar\sigma} +
\big( n_{\bar\sigma} -2\big)
n_{\bar\sigma} {\sf n}_{\sigma}^{(0)}({\Bf k}) \nonumber\\
&&\;\;\;\;
-\frac{1}{N_{\sc l}^2} \sum_{{\bF k}_1,{\bF k}_2}
{\sf n}_{\bar\sigma}^{(0)}({\Bf k}_1)
{\sf n}_{\bar\sigma}^{(0)}({\Bf k}_2)
{\sf n}_{\sigma}^{(0)}({\Bf k}-{\Bf k}_1+{\Bf k}_2).
\end{eqnarray}
From the result in Eq.~(\ref{ea17}), one readily obtains
\begin{equation}
\label{ea18}
\frac{1}{N_{\sc l}} \sum_{\bF k} 
\delta {\sf n}_{\sigma}^{(0)}({\Bf k}) = 
- n_{\sigma} n_{\bar\sigma} \le 0, 
\end{equation}
in conformity with the exact result in Eq.~(\ref{ea12}) above. 
The deviation of the RHS of Eq.~(\ref{ea18}) from zero is
a measure for the violation of ${\hat c}_{i;\bar\sigma} 
{\hat c}_{i;\sigma} \vert\Psi_{N;0}\rangle = 0$, $\forall i$, 
for $\vert\Psi_{N;0}\rangle$ replaced by $\vert\Phi_{N;0}\rangle$
(see text following Eq.~(\ref{ea10})).
\begin{figure}[t!]
\protect
\centerline{
\psfig{figure=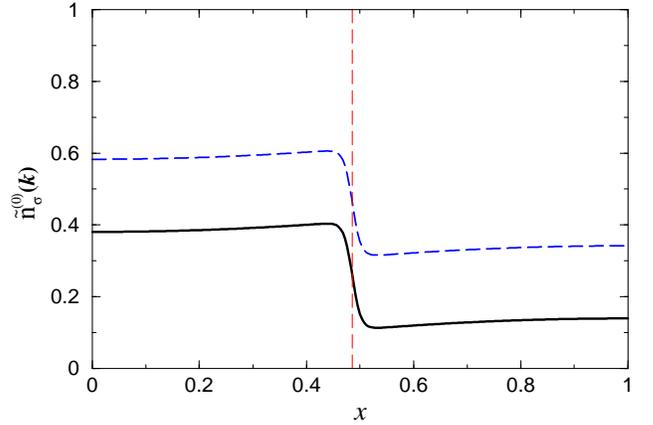,width=3.25in} }
\vskip 5pt
\caption{\label{fia1} \sf
The momentum distribution function $\tilde{\sf n}_{\sigma}^{(0)}
({\Bf k})$ (solid line) pertaining to a square lattice for $d=2$ for 
${\Bf k} = x (\pi,\pi)$ (in units of the inverse lattice constant)
corresponding to $n_{\sigma} = n_{\bar\sigma} = 9/20$, at temperature 
$T=0.1 t/k_{\sc b}$. The curve for $\tilde{\sf n}_{\sigma}^{(0)}
({\Bf k}) + n_{\sigma} n_{\bar\sigma}$ (broken line) should be 
compared with curves in Fig.~1 of \protect\cite{PLS98a}.
We have taken account of $T$ by replacing ${\sf n}_{\sigma}^{(0)}
({\Bf k})$, $\forall {\Bf k},\sigma$, on the RHS of
Eq.~(\protect\ref{ea17}) by the non-interacting Fermi function
introduced in Eq.~(\protect\ref{ea19}). For completeness, here 
${\Bf k}_{{\sc f};\sigma}$ corresponds to $x_{{\sc f};\sigma} = 
0.485686\dots$ (indicated by the vertical broken line); at 
$x_{{\sc f};\sigma}$, $\tilde{\sf n}_{\sigma}^{(0)}({\Bf k}) + 
n_{\sigma} n_{\bar\sigma}$ is approximately equal to $0.461$, that 
is slightly less than $1/2$. Further, in the vicinity of $x=0.436$,
$\tilde{\sf n}_{\sigma}^{(0)}({\Bf k}) + n_{\sigma} n_{\bar\sigma}
\approx 0.4036 + 0.2025 = 0.6061 > 1 - n_{\sigma} = 0.55$, in
violation of the inequality in Eq.~(\protect\ref{ea24}). }
\end{figure}

In Fig.~\ref{fia1} we depict $\tilde{\sf n}_{\sigma}^{(0)}({\Bf k})$ as 
expressed in Eq.~(\ref{ea17}) above in which we have, however, 
replaced all ${\sf n}_{\sigma}^{(0)}({\Bf k})$, 
$\forall {\Bf k},\sigma$, by the non-interacting Fermi function 
\begin{equation}
\label{ea19}
f_{\sigma}({\Bf k}) {:=} 
\frac{1}{1+\exp([\epsilon_{{\bF k};\sigma} - 
\varepsilon_{\sc f}^{(0)}]/[k_{\sc b} T])}.
\end{equation}
The function depicted in Fig.~\ref{fia1} concerns ${\Bf k}$ along 
the diagonal direction of the 1BZ pertaining to a square lattice 
and corresponds to $n_{\sigma} = n_{\bar\sigma} = 9/20$ and 
$T = 0.1 t/k_{\sc b}$. This Figure is therefore to be compared 
with Fig.~1 of \cite{PLS98a}. We should point out that 
$\tilde{\sf n}_{\sigma}({\Bf k})$ in both \cite{PLS98a} and 
\cite{PLS98b} (corresponding to $\vert\Psi_{N;0}\rangle$) are 
subject to the normalization condition $N_{\sc l}^{-1} \sum_{\bF k} 
\tilde{\sf n}_{\sigma}({\Bf k}) = n_{\sigma}$ ({\it cf}. 
Eq.~(\ref{ea13}) and the subsequent remark). 

By attributing $-n_{\sigma} n_{\bar\sigma}$ (which in the present 
case is equal to $-0.2025$) on the RHS of Eq.~(\ref{ea18}) to a 
uniform underestimation of $\tilde{\sf n}_{\sigma}({\Bf k})$ by 
$\tilde{\sf n}_{\sigma}^{(0)}({\Bf k})$ over the 1BZ (see
Eq.~(\ref{ea9})), $\tilde{\sf n}_{\sigma}^{(0)}({\Bf k})+n_{\sigma} 
n_{\bar\sigma}$, also depicted in Fig.~\ref{fia1}, would be the 
function to be directly compared with the functions depicted in 
Fig.~1 of \cite{PLS98a}. Indeed, the upward shift of 
$\tilde{\sf n}_{\sigma}^{(0)}({\Bf k})$ by $n_{\sigma} n_{\bar\sigma}$ 
brings the resulting function into relatively good quantitative 
agreement with the results in \cite{PLS98a}; the results are 
also comparable with those as calculated by Stephan and Horsch 
\cite{WSPH91} (see Table~I herein; for an explicit comparison of 
the latter results with those based on an high-temperature expansion, 
see Fig.~2 in \cite{SG92}); for instance, for $J/t=0.4$ and two holes 
per 20 sites (i.e. for $n=0.9$, or $n_{\sigma}=n_{\bar\sigma}=0.45$) 
the latter workers obtained $0.54681$ and $0.36247$ for ${\Bf k}$ at 
$(0,0)$ and $(\pi,\pi)$ (in units of the inverse lattice constant)
respectively, to be compared with our upward-shifted results 
which for the indicated ${\Bf k}$ points are equal to $0.583$ and 
$0.342$ respectively. We should point out that the uniform increase in
$\tilde{\sf n}_{\sigma}^{(0)}({\Bf k})$ by $n_{\sigma} n_{\bar\sigma}$ 
(in order to account approximately for the consequence of the 
prohibition of double occupancy in the correlated state 
$\vert\Psi_{N;0}\rangle$) may at some regions of the 1BZ give rise 
to violation of an exact result (see Eq.~(\ref{ea24}) below); 
according to this result, in the present case $\tilde{\sf n}_{\sigma}
({\Bf k})$ is bound to satisfy $\tilde{\sf n}_{\sigma}({\Bf k}) \le 
1-0.45=0.55$.

For completeness, $\tilde{\sf n}_{\sigma}({\Bf k})$ can be associated 
with a Green function $\wt{G}_{\sigma}'({\Bf r},{\Bf r}';z)$ similar 
to $\wt{G}_{\sigma}({\Bf r},{\Bf r}';z)$ in Eq.~(\ref{e3}) above 
whose corresponding Lehmann amplitudes are defined exactly as in 
Eq.~(\ref{e5}), however, in terms of ${\hat a}_{{\bF k};\sigma}$. It 
follows that for the single-particle spectral function $A_{\sigma}'
({\Bf k};\varepsilon)$ associated with $\wt{G}_{\sigma}'({\Bf r},
{\Bf r}';z)$ ({\it cf}. Eq.~(\ref{e9})) we have
\begin{equation}
\label{ea20}
\frac{1}{\hbar} \int_{-\infty}^{\infty} {\rm d}\varepsilon\;
A_{\sigma}'({\Bf k};\varepsilon) 
= \langle\Upsilon_{N}\vert [ {\hat a}_{{\bF k};\sigma}^{\dag},
{\hat a}_{{\bF k};\sigma}]_{+} \vert\Upsilon_{N}\rangle.
\end{equation}
This is exactly the result applicable to $A_{\sigma}({\Bf k};
\varepsilon)$, with the exception that, for the latter function,
$\big[ {\hat a}_{{\bF k};\sigma}^{\dag}, {\hat a}_{{\bF k};\sigma} 
\big]_{+}$ on the RHS of Eq.~(\ref{ea20}) has to be replaced by 
$[{\hat c}_{{\bF k};\sigma}^{\dag}, {\hat c}_{{\bF k};\sigma}]_{+} 
=1$, hence we have Eqs.~(\ref{e10}) and (\ref{e11}), 
according to which $\hbar^{-1} \int_{-\infty}^{\infty} 
{\rm d}\varepsilon\, A_{\sigma}({\Bf k};\varepsilon) = 1$. In the 
present case, where
\begin{equation}
\label{ea21}
\big[ {\hat a}_{i;\sigma}^{\dag}, {\hat a}_{j;\sigma} \big]_{+} 
= (1 - {\hat n}_{i;\bar\sigma}) \delta_{i,j},
\end{equation} 
for uniform GSs, Eq.~(\ref{ea20}) can be written as
\begin{equation}
\label{ea22}
\frac{1}{\hbar} \int_{-\infty}^{\infty} {\rm d}\varepsilon\;
A_{\sigma}'({\Bf k};\varepsilon)
= 1 - n_{\bar\sigma}.
\end{equation}
This result has been earlier reported by Stephan and Horsch 
\cite{WSPH91}. Since $A_{\sigma}'({\Bf k};\varepsilon)$, exactly
as $A_{\sigma}({\Bf k};\varepsilon)$, is positive semi definite, 
it follows that 
\begin{equation}
\label{ea23}
\frac{1}{\hbar} \int_{-\infty}^{\varepsilon_0} {\rm d}\varepsilon\;
A_{\sigma}'({\Bf k};\varepsilon) \le 1 - n_{\bar\sigma}\;\;\;
\mbox{\rm for any}\;\; \varepsilon_0.
\end{equation}
With $\mu'$ the counterpart of $\mu$ in Eqs.~(\ref{e5})-(\ref{e8}),
we thus have
\begin{equation}
\label{ea24}
\tilde{\sf n}_{\sigma}({\Bf k}) \equiv
\frac{1}{\hbar} \int_{-\infty}^{\mu'} {\rm d}\varepsilon\;
A_{\sigma}'({\Bf k};\varepsilon) \le 1 - n_{\bar\sigma}.
\end{equation}
This inequality has been invoked by Putikka, {\sl et al.} 
\cite{PLS98a}, adding support to the reliability of their 
numerical results. We note in passing that the maximum value 
taken by $\tilde{\sf n}_{\sigma}^{(0)}({\Bf k})$ in Fig.~\ref{fia1}
is equal to $0.4036$ (corresponding to $x=0.436$) which is indeed 
less than $1-9/20=0.55$. As we have indicated above (see our 
considerations following Eq.~(\ref{ea19})), in the present 
case, where $n_{\sigma} n_{\bar\sigma} = 0.2025$, a uniform increase 
by $n_{\sigma} n_{\bar\sigma}$ of $\tilde{\sf n}_{\sigma}^{(0)}
({\Bf k})$ over the 1BZ gives rise to the violation of the 
inequality in Eq.~(\ref{ea24}).
\hfill $\Box$

\end{appendix}



\end{document}